\documentclass[12pt]{article}
\usepackage[a4paper, margin=1in]{geometry}
\usepackage{amsmath, amssymb}
\usepackage{authblk}
\usepackage{graphicx}
\usepackage{cite}
\usepackage{hyperref}
\usepackage{subcaption}
\usepackage{float}
\usepackage[section]{placeins}
\usepackage{epstopdf}

\title{\bf A Combined Barrow Entropy and QCD Ghost Mechanism for Late-Time Cosmic Acceleration
}

\author[1]{Aziza Altaibayeva\thanks{aziza.ltaibayeva@gmail.com 
}}
\author[2]{Ulbossyn Ualikhanova\thanks{ulbossyn.ualikhanova@gmail.com 
}}
\author[3]{Zhanar Umurzakhova\thanks{zhumurzakhova@gmail.com
}}
\affil[1,2,3]{\small Department of General and Theoretical Physics, L.N. Gumilyov Eurasian National University, Astana 010008,Kazakhstan.}
\author[3]{Surajit Chattopadhyay \footnote{Corresponding author}\thanks{schattopadhyay1@kol.amity.edu; surajitchatto@outlook.com
}}
\affil[3]{Department of Mathematics, Amity University Kolkata, Major Arterial Road, Action Area II, Rajarhat, Newtown, Kolkata 700135, India }

\date{\today}

\begin{document}

\maketitle

\begin{abstract}
We investigate a unified dark-energy scenario based on the combined effects of Barrow entropy corrections and the QCD ghost mechanism, referred to as the BH--QCDGDE model. The dark-energy density is constructed in a generalized holographic form that incorporates both Barrow-deformed entropy corrections and low-energy QCD vacuum effects within a single framework. The cosmological dynamics are analyzed in a spatially flat Friedmann--Lema\^{\i}tre--Robertson--Walker
background. The model exhibits a smooth transition from a decelerated matter-dominated era to a late-time accelerated phase without crossing the phantom divide, indicating a viable background evolution. An equivalent scalar-field description of the effective dark-energy sector is reconstructed and shown to admit a quintessence-like behavior. The thermodynamic viability is examined by testing the generalized second law at the apparent horizon, which is found to be satisfied throughout the parameter space. The classical stability of the model is further investigated through the squared speed of sound, revealing
the role of model parameters in shaping stable cosmological regimes. Overall, the BH--QCDGDE framework provides a consistent and physically viable description of late-time cosmic acceleration.
\end{abstract}

\section{Introduction}
The observed accelerated expansion of the universe, inferred from Type Ia Supernovae \cite{Riess1998,Perlmutter1999} and supported by Cosmic Microwave Background (CMB) and Baryon Acoustic Oscillation (BAO) data \cite{Planck2020}, remains one of the central puzzles in modern cosmology. The standard $\Lambda$CDM model, though phenomenologically successful, is plagued by fine-tuning and coincidence problems \cite{Weinberg1989,Peebles2003}. The term ``dark energy" began to appear in the scientific works in 1998, following the discovery of the accelerated expansion of the Universe \cite{Riess1998,Perlmutter1999,Perlmutter1998} announced almost simultaneously by the Supernova Cosmology Project and the High-Z Supernova Search Team. However, the concept itself predates this terminology \cite{Novosyadlyj2013}. In particular, Turner \cite{Turner2001} referred to this unknown component as a smooth component, while Paul Steinhardt \cite{Steinhardt2000} introduced the term quintessence to describe a dynamical form of cosmic acceleration. Consequently, several dynamical dark energy (DE) and modified gravity scenarios have been explored \cite{Copeland2006,Nojiri2011,Saridakis2021}. 

Dark energy encompasses a broad spectrum of hypothetical physical entities proposed to account for the late-time accelerated expansion of the Universe \cite{Brax2018}. Notably, the search for such a component has long been an integral pursuit in cosmology, predating its observational confirmation \cite{Novosyadlyj2013}. Among them, the holographic dark energy (HDE) framework connects cosmic evolution with horizon entropy \cite{Cohen1999,Li2004}. Generalized holographic dark energy (GHDE) has emerged as a unifying framework capable of incorporating a wide class of infrared cutoffs and entropy corrections within a covariant formulation. In this context, Nojiri and Odintsov first proposed a covariant generalized holographic dark energy model that naturally realizes late-time cosmic acceleration without violating fundamental principles of general relativity \cite{Nojiri2017}. This framework was subsequently extended to provide a unified description of early-time inflation and late-time acceleration within a single holographic setup \cite{Nojiri2020}. The versatility of GHDE models, encompassing various phenomenological realizations and cosmological behaviors, was further systematized in \cite{Nojiri2021}. Extensions involving dissipative effects, such as viscosity, have also been explored using the Nojiri--Odintsov cutoff, leading to rich cosmic dynamics \cite{Khurshudyan2016}. More recently, Barrow entropic dark energy has been shown to arise as a specific member of the GHDE family, highlighting the deep connection between quantum-gravitational entropy corrections and holographic cosmology \cite{Nojiri2022}. Earlier investigations that are involved in linking vacuum fluctuations, holography, and effective phantom behavior further reinforce the foundational role of holographic principles in dark-energy physics \cite{Elizalde2005}. The present work fits naturally within this generalized holographic paradigm by incorporating Barrow entropy deformation together with QCD ghost contributions, thereby extending the GHDE framework to a broader and physically motivated dark-energy scenario.

Recently, Barrow \cite{Barrow2020} proposed a modified entropy--area relation to incorporate quantum gravitational effects, leading to Barrow holographic dark energy (BHDE) models \cite{Saridakis2020}. In parallel, the QCD ghost dark energy (QCDGDE) mechanism arises from the non-trivial contribution of the Veneziano ghost field in quantum chromodynamics, which yields a vacuum energy density proportional to the Hubble parameter $H$ \cite{Urban2009,Ohta2011}. Motivated by the complementary nature of these two approaches, we construct a new hybrid model, the the Barrow Holographic QCD Ghost Dark Energy (BH--QCDGDE), which unifies the entropy-modified holographic term with the ghost-induced vacuum contribution. This framework allows one to explore the interplay between quantum gravitational corrections and QCD vacuum effects in driving late-time cosmic acceleration. At this juncture, it is worth noting that the current work is a substantial continuation of our previous study, published in \cite{Altaibayeva2024PhysScr}, where we primarily investigated the cosmology of Barrow holographic QCD ghost dark energy by reconstructing the Hubble parameter and analyzing the background dynamics and thermodynamics at the apparent horizon. Unlike that work~\cite{Altaibayeva2024PhysScr}, the present analysis takes a fully generalized formulation of the dark-energy density, explicitly including QCD ghost contributions and Barrow entropy deformation in a single BH--QCDGDE framework. 

The organization of the paper is as follows. In Sec.~2, we formulate the BH--QCDGDE model by constructing the generalized dark-energy density and derive the modified Friedmann equations governing the background cosmological dynamics. The evolution of the Hubble parameter, the total equation-of-state parameter, and the deceleration parameter are analyzed in detail, along with their three-dimensional parameter-space
behavior. In Sec.~3, we reconstruct an equivalent scalar-field description of the effective dark-energy sector and investigate the associated dynamical properties. Section~4 is devoted to an analytical study based on a power-law scale factor, where explicit expressions for the energy density, equation of state, scalar field, and potential are obtained. In Sec.~5, we reconstruct the scale factor by considering a perturbative Barrow deformation and examine the resulting late-time cosmological behavior. The validity of the generalized second law of thermodynamics is tested at the apparent horizon in Sec.~6. In Sec.~7, we analyze the classical stability of the effective cosmic fluid through the squared speed of sound and identify the stable regions of the parameter space. Finally, Sec.~8 summarizes the main results and discusses possible directions for future investigations.

\section{The BH--QCDGDE Model Formulation}
\subsection{Modified dark energy density}
In the present subsection of this work, we consider a generalized form of the dark energy density that is motivated by recent developments in both holographic and quantum chromodynamic (QCD) sectors ~\cite{Yang2022}. The idea was generated from the growing evidence from the existing literature that quantum gravitational effects and vacuum contributions from QCD may jointly influence the large-scale dynamics of the universe ~\cite{Kovchegov2013,Brodsky2011,Armesto2014}. In particular, the deformation of the horizon entropy, as proposed by Barrow, modifies the conventional holographic dark energy behaviour, while the Veneziano ghost field in QCD introduces a dynamical vacuum energy component proportional to the Hubble parameter \cite{Barrow2020,Saridakis2020,Dabrowski2020,Sheykhi2023,Das2023,Luciano2025EPJC,Luciano2025arxiv}. 

From the combination of these two effects, we now construct a modified dark energy density that incorporates the essential features of both the Barrow holographic and the QCD ghost contributions. The total dark energy density is taken as
\begin{equation}
\rho_{\mathrm{DE}}(H) = 3M_{\mathrm{Pl}}^2 \left( \alpha H^{2-\Delta_B} + \beta H \right),
\label{rhoDE}
\end{equation}
where $\Delta_B$ denotes the Barrow deformation parameter ($0 \le \Delta_B \le 1$) \cite{Oliveros2022}, $\alpha$ controls the Barrow holographic term, and $\beta$ scales the QCD ghost contribution. The proposed dark energy density is inspired by two independent but complementary theoretical considerations, namely the Barrow holographic principle  \cite{Barrow2020, Saridakis2022} and the QCD ghost dark energy hypothesis ~\cite{Urban2009,Ohta2011,Cai2012}. Their combination provides a unified phenomenological framework for explaining the late-time cosmic acceleration without invoking any additional scalar field or cosmological constant.

The form of the dark energy density adopted in Eq.~\eqref{rhoDE},
is motivated by the consideration that the late--time acceleration of the Universe may arise from the combined effects of quantum--gravitational corrections to horizon entropy and vacuum contributions originating from quantum chromodynamics (QCD) with some motivation from the work of \cite{Luciano2025EPJC}. Independently, the QCD ghost dark energy mechanism predicts a dynamical vacuum energy density proportional to the Hubble parameter, $\rho \propto H$, arising from the
Veneziano ghost contribution in an expanding spacetime \cite{Cai2012Ghost}. Although the fact that the ghost field is unphysical in Minkowski spacetime, it generates a non-vanishing vacuum energy in a time-dependent background of cosmologyd, which naturally results in the correct order of magnitude for the observed dark energy density \cite{Urban2009}.

The starting point lies in the holographic principle ~\cite{tHooft2001,Bousso2000,Bousso2002}, which asserts that the number of physical degrees of freedom in a volume of space is determined by its boundary area \cite{Cohen1999}. In the cosmological context, the total vacuum energy enclosed within a region of characteristic length scale $L$ must satisfy
\begin{equation}
L^3 \rho_{\mathrm{DE}} \leq M_{\mathrm{Pl}}^2 L,
\end{equation}
which, when saturated, leads to the standard holographic dark energy (HDE) density
\begin{equation}
\rho_{\mathrm{HDE}} = 3 c^2 M_{\mathrm{Pl}}^2 L^{-2}.
\end{equation}
If one adopts the Hubble horizon ($L = H^{-1}$) as the infrared (IR) cutoff, the energy density scales as $\rho_{\mathrm{HDE}} \propto H^2$.

Barrow \cite{Barrow2020} suggested that quantum gravitational effects could render the geometry of the black-hole horizon fractal-like, leading to a modified entropy--area relation of the form
\begin{equation}
S_{B} = \left( \frac{A}{A_0} \right)^{1+\frac{\Delta_B}{2}},
\end{equation}
where $\Delta_B$ is the Barrow exponent ($0 \leq \Delta_B \leq 1$). The standard Bekenstein--Hawking entropy is recovered for $\Delta_B = 0$, while $\Delta_B = 1$ corresponds to the maximally deformed case. Incorporating this entropy deformation into the holographic framework modifies the energy density to
\begin{equation}
\rho_{\mathrm{BHDE}} \propto H^{2 - \Delta_B},
\end{equation}
thus introducing the first term $\alpha H^{2-\Delta_B}$ in Eq.~\eqref{rhoDE}. This term effectively encodes the quantum gravitational corrections to the dark energy density through the fractal deformation parameter $\Delta_B$.

The second term in Eq.~\eqref{rhoDE}, proportional to $H$, originates from the QCD ghost dark energy (QGDE) framework \cite{Urban2009,Ohta2011}. In quantum chromodynamics, the Veneziano ghost field, which resolves the $U\eqref{rhoDE}_A$ problem, contributes to the vacuum energy density in a time-dependent background. Although this field is unphysical in Minkowski spacetime, in an expanding universe it induces a small vacuum energy density of the form
\begin{equation}
\rho_{\mathrm{QGDE}} \propto H.
\end{equation}
This contribution naturally yields the correct order of magnitude for the observed dark energy density without fine-tuning, and it evolves mildly with cosmic time.

By combining these two effects, we arrive at the expression in Eq.~\eqref{rhoDE}, which unifies the Barrow holographic modification ($\propto H^{2-\Delta_B}$) with the QCD ghost term ($\propto H$). The constants $\alpha$ and $\beta$ quantify the relative strengths of the two contributions, while the parameter $\Delta_B$ measures the deviation from the standard holographic entropy law. 

Physically, this hybrid form allows for a rich dynamical behaviour of the dark energy sector. At early times, the holographic term dominates, while at late times, the QCD ghost term drives the accelerated expansion. The model therefore provides a smooth transition from a matter-dominated universe to a dark-energy-dominated era, consistent with current cosmological observations.

Such a construction not only connects quantum gravity corrections (through $\Delta_B$) and low-energy QCD effects (through $\beta$), but also offers a theoretically well-motivated and observationally viable alternative to the cosmological constant for explaining the late-time acceleration of the universe

Substituting Eqs.~(\ref{rhoDE}) into the conservation equation and solving for the equation-of-state parameter, we find
\begin{equation}
w_{\rm DE}
=-1
-\frac{\left[\alpha(2-\Delta_B)H^{1-\Delta_B}+\beta\right]\dot{H}}
{H\left(\alpha H^{2-\Delta_B}+\beta H\right)}.
\label{wDE}
\end{equation}

From \eqref{wDE} it is understandable that the dark energy density represented by \eqref{rhoDE} can lead to quientessence or phantom according as $\frac{\left[\alpha(2-\Delta_B)H^{1-\Delta_B}+\beta\right]\dot{H}}
{H\left(\alpha H^{2-\Delta_B}+\beta H\right)}<~\text{or}~>0$. At this juncture, let us try to re-express \eqref{wDE} purely in terms of redshift $z$. For that purpose, first we utilize the redshift--time relation
\begin{equation}
\dot{H}=-(1+z)H(z)\frac{dH(z)}{dz},
\label{Hdotz}
\end{equation}
and the dark energy density given in Eq.~(\ref{rhoDE}),
the equation-of-state parameter can be written purely as a function of the
redshift:
\begin{equation}
w_{\rm DE}(z)
=
-1
+
\frac{(1+z)\left[\alpha(2-\Delta_B)H^{1-\Delta_B}(z)+\beta\right]}
{\alpha H^{2-\Delta_B}(z)+\beta H(z)}
\frac{dH(z)}{dz}.
\label{wDEz}
\end{equation}

Introducing the dimensionless Hubble parameter
$E(z)=H(z)/H_0$, Eq.~(\ref{wDEz}) takes the compact form
\begin{equation}
w_{\rm DE}(z)
=-1+\frac{(1+z)\left[\alpha(2-\Delta_B)E^{1-\Delta_B}(z)+\beta\right]}
{\alpha E^{2-\Delta_B}(z)+\beta E(z)}
\frac{dE(z)}{dz}.
\label{wDEEz}
\end{equation} 

For an expanding Universe, the Hubble parameter decreases monotonically with cosmic time, and hence $\dot{H}<0$ implying 
$dE/dz>0$. In \eqref{wDEEz} we have $\alpha E^{2-\Delta_B}(z)+\beta E(z)>0$. Thus, phantom or quintessence would depend on $\alpha(2-\Delta_B)E^{1-\Delta_B}(z)+\beta$. For positive values of the model parameters $\alpha$ and $\beta$, the
correction term in Eq.~(\ref{wDEEz}) is therefore positive. As a result, the BH--QCDGDE framework realizes quintessence regime, with $w_{\rm DE}$ approaching $-1$ asymptotically from above. 

\subsection{Background cosmological dynamics}
To examine the cosmological implications of the proposed BH--QCDGDE framework, we will now investigate the background dynamics in a spatially flat FRW universe. By incorporating the modified dark energy density into the Friedmann equations, we would now attempt to derive the evolutionary behaviour of the Hubble parameter and analyze how the interplay between the Barrow entropy deformation and the QCD ghost contribution influences the expansion of the universe. For a spatially flat FRW universe, the first Friedmann equation reads
\begin{equation}
3 M_{\mathrm{Pl}}^2 H^2 = \rho_m + \rho_r + \rho_{\mathrm{DE}}(H),
\label{friedmann}
\end{equation}
where $H$ is the Hubble parameter, $M_{\mathrm{Pl}}$ is the reduced Planck mass, and $\rho_m$ and $\rho_r$ are the energy densities of pressureless matter and radiation, respectively. The dark energy component $\rho_{\mathrm{DE}}(H)$ corresponds to the Barrow holographic dark energy (BHDE), which depends explicitly on the Hubble parameter through the quantum-gravitational deformation exponent $\Delta_B$. Standard energy conservation equations for matter and radiation imply $\rho_m \propto (1+z)^3$ and $\rho_r \propto (1+z)^4$, where $z$ is the cosmological redshift. Introducing the dimensionless Hubble parameter
\begin{equation}
E(z) \equiv \frac{H(z)}{H_0},
\end{equation}
with $H_0$ being the present-day Hubble constant, and the current density parameters
\begin{equation}
\Omega_{m0} \equiv \frac{\rho_{m0}}{3 M_{\mathrm{Pl}}^2 H_0^2}, \quad 
\Omega_{r0} \equiv \frac{\rho_{r0}}{3 M_{\mathrm{Pl}}^2 H_0^2},
\end{equation}
the Friedmann equation \eqref{friedmann} can be recast in the form
\begin{equation}
E^2(z) = \Omega_{m0} (1+z)^3 + \Omega_{r0} (1+z)^4 + \alpha E^{2-\Delta_B}(z) + \beta E(z),
\label{Ez}
\end{equation}
where $\alpha$ and $\beta$ are dimensionless parameters characterizing the BHDE contribution. The first term in the BHDE density, proportional to $E^{2-\Delta_B}$, arises from the Barrow entropy deformation, while the linear term in $E(z)$ provides additional phenomenological flexibility.  
Hence we have
\begin{equation}
\frac{dE}{dz} = 
\frac{3 \, \Omega_{m0} (1+z)^2 + 4 \, \Omega_{r0} (1+z)^3}
{\,2 E(z) - \alpha (2-\Delta_B)\, E^{1-\Delta_B}(z) - \beta\,}.
\label{eq:dEdz}
\end{equation}

Numerical integration of Eq.~\eqref{Ez} reveals a smooth transition from an early matter-dominated epoch to a late-time accelerating phase. The present values $q_0 \approx -0.54$ and $\omega_{\mathrm{tot},0} \approx -0.69$ are consistent with an accelerated universe without crossing the phantom divide ($\omega_{\mathrm{tot}} > -1$).  
The transition redshift is found to be $z_{\mathrm{tr}} \approx 0.65$, implying a gradual shift to acceleration driven by the hybrid dark energy density \eqref{rhoDE}.

Here, we consider Eq. \eqref{eq:dEdz}, which is a first-order nonlinear differential equation.  We use the method of
separation of variables to solve this equation. For that purpose, we first rearrange the terms of \eqref{eq:dEdz} as follows:

\begin{equation}
\left[2E - \alpha(2-\Delta_B)E^{1-\Delta_B} - \beta\right] dE
=
\left[3\Omega_{m0}(1+z)^2 + 4\Omega_{r0}(1+z)^3\right] dz .
\label{eq:separated}
\end{equation}

On integration of both sides of \eqref{eq:separated}, the general implicit solution of \eqref{eq:dEdz} comes out to be
\begin{equation}
E^2(z) - \beta E(z) - \alpha E^{2-\Delta_B}(z)
=
\Omega_{m0}(1+z)^3 + \Omega_{r0}(1+z)^4 + C .
\label{eq:implicit}
\end{equation}

Considering $z=0$, we have $C = 1 - \Omega_{m0} - \Omega_{r0} - \alpha - \beta$, which can be considered for the particular solution. Now, we consider three different cases of $\Delta_B$. First we consider $\Delta_B=0$. In this case, Eq.~\eqref{eq:implicit} reduces to a quadratic equation in $E(z)$ and solution is 

\begin{equation}
E(z)
=
\frac{
\beta
+
\sqrt{
\beta^{2}
+
4(1-\alpha)
\left[
\Omega_{m0}(1+z)^{3}
+
\Omega_{r0}(1+z)^{4}
+
1-\Omega_{m0}-\Omega_{r0}-\alpha-\beta
\right]
}
}{
2(1-\alpha)
}.
\label{eq:E_delta0}
\end{equation}
This case corresponds to the standard QCD ghost dark energy scenario
with a quadratic holographic contribution.

Next we consider $\Delta_B=1$, where we have

\begin{equation}
E(z) =
\frac{(\alpha+\beta)
+\sqrt{
(\alpha+\beta)^{2}
+
4\left[
\Omega_{m0}(1+z)^{3}
+ \Omega_{r0}(1+z)^{4}
+ 1-\Omega_{m0}-\Omega_{r0}-\alpha-\beta
\right]
}
}{2}.
\label{eq:E_delta1}
\end{equation}

This case represents the maximally deformed Barrow entropy scenario \cite{SaridakisBasilakos2021,Salehi2023}, where the holographic term scales linearly with $H$.

For $\Delta_B=2$, the corresponding solution is
\begin{equation}
E(z)
=
\frac{
\beta
+
\sqrt{
\beta^{2}
+
4\left[
\Omega_{m0}(1+z)^{3}
+
\Omega_{r0}(1+z)^{4}
+
1-\Omega_{m0}-\Omega_{r0}-\beta
\right]
}
}{2}.
\label{eq:E_delta2}
\end{equation}

The $\Delta=2$ is not part of standard Barrow holographic dark energy. In this case, the Barrow contribution behaves as a constant vacuum term,
effectively mimicking a cosmological-constant–like correction \cite{Sharma2022}.

The total equation of state (EoS) parameter and the deceleration parameter are defined as \cite{Wei2007,Zhang2008,DelCampo2011}
\begin{align}
w_{\mathrm{tot}}(z) &= -1 + \frac{2(1+z)}{3 E(z)} \frac{dE(z)}{dz},\label{w}\\
q(z) &= -1 + (1+z) \frac{1}{E(z)} \frac{dE(z)}{dz},\label{q}.
\end{align}
A positive deceleration parameter $q>0$ corresponds to a decelerated expansion, while $q<0$ indicates acceleration. The transition redshift $z_{\mathrm{tr}}$, defined implicitly through
\begin{equation}
q(z_{\mathrm{tr}}) = 0,
\end{equation}
marks the epoch at which the Universe transitions from deceleration to acceleration, providing a key observational probe of the cosmic expansion history.

\begin{figure}[htbp]
    \centering
    \includegraphics[width=0.75\textwidth]{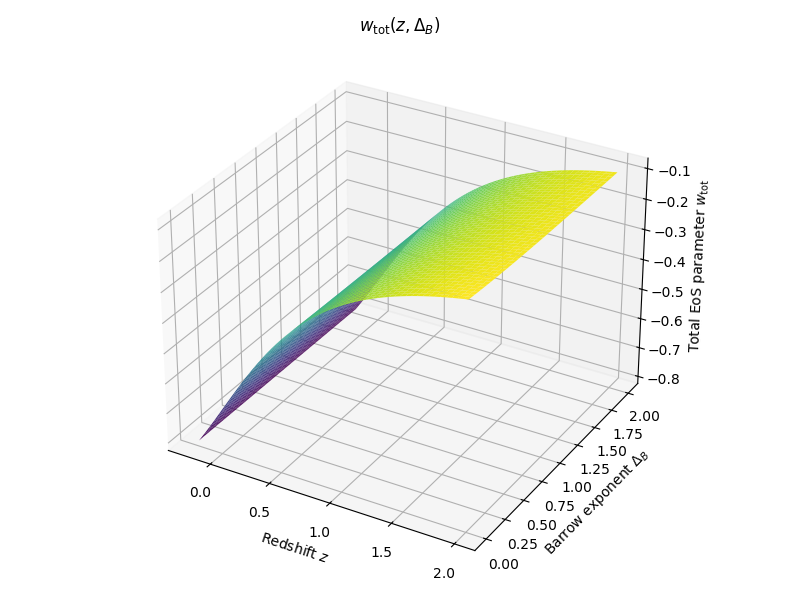}
    \caption{
    Three-dimensional evolution of the total equation-of-state parameter~$w_{\rm tot}(z,\Delta)$ for the BH--QCDGDE model as a function of the redshift $z$ and the Barrow exponent $\Delta$. The three-dimensional plot is constructed using Eq.~\eqref{w}, where the Hubble function $E(z)$ is constrained by Eq.~\eqref{eq:implicit} and its redshift derivative is determined from Eq.~\eqref{eq:dEdz}. The parameter values are taken as $\Omega_{m0}=0.3$, $\Omega_{r0}=8\times10^{-5}$, $\alpha=0.02$, and $\beta=0.08$.
    }
    \label{fig:wtot_3D}
\end{figure}

\begin{figure}[htbp]
    \centering
    \includegraphics[width=0.7\textwidth]{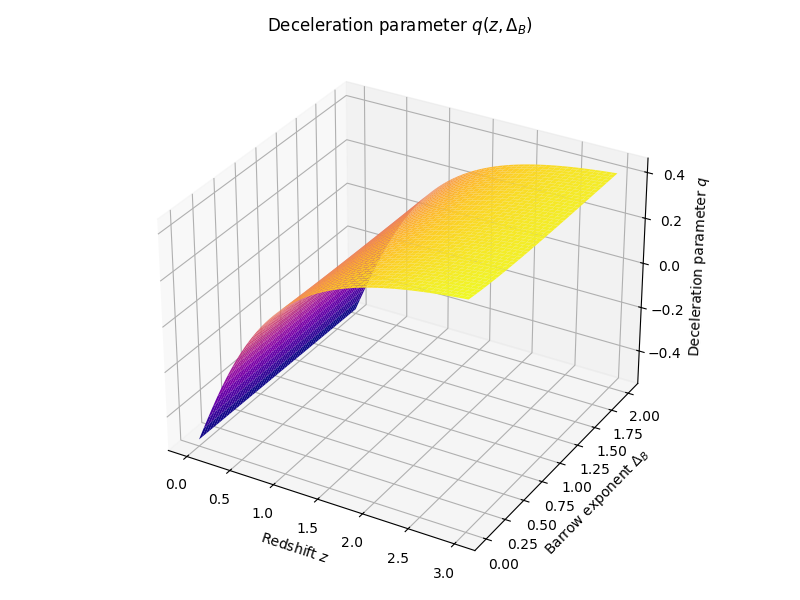}
    \caption{Three-dimensional evolution of the deceleration parameter
    $q(z,\Delta_B)$ as a function of the redshift $z$ and the Barrow exponent
    $\Delta_B$, obtained using Eq.~\eqref{q}. The dimensionless Hubble function
    $E(z)$ is constrained by the implicit solution in Eq.~\eqref{eq:implicit}, while its
    redshift derivative is determined from Eq.~\eqref{eq:dEdz}. The surface clearly
    illustrates the transition from a decelerated expansion phase ($q>0$) at
    higher redshifts to a late-time accelerated phase ($q<0$), with the Barrow
    entropy deformation controlling the smoothness of the transition.}
    \label{fig:q_DeltaB}
\end{figure}

The three-dimensional evolution of the total equation-of-state parameter, as presented in Fig. \ref{fig:wtot_3D}, the 
$w_{\rm tot}(z,\Delta)$ demonstrates that the BH--QCDGDE model is in alignment with late-time accelerated expansion, with $w_{\rm tot}(0)<-1/3$. The surface that the cosmological dynamics is not significantly sensitive to the specific value of the Barrow exponent.
The dominant evolution of $w_{\rm tot}$ is governed by the redshift dependence, with a smooth transition from a matter-dominated phase at higher redshifts to a dark-energy–dominated accelerating phase at late times. Furthermore, the absence of phantom crossing in the future evolution suggests a stable cosmological scenario. The three-dimensional evolutionary pattern of the deceleration parameter $q(z,\Delta_B)$ as a function of the redshift $z$ and the Barrow exponent $\Delta_B$ is presented in Fig.~\ref{fig:q_DeltaB}. This figure reveals a smooth transition from a decelerated expansion phase ($q>0$) at higher redshifts to an accelerated phase ($q<0$) at late times. The significant contribution to the evolution of $q$ results from the redshift. It is also observed from Fig.~\ref{fig:q_DeltaB} that variations in $\Delta_B$ do not bring in any noticeable modulation of the transition epoch. This behavior indicates that the Barrow entropy deformation provides a physically consistent description of the late-time acceleration of the Universe.

\begin{figure}[htbp]
    \centering

    \begin{subfigure}[b]{0.32\textwidth}
        \centering
        \includegraphics[width=\textwidth]{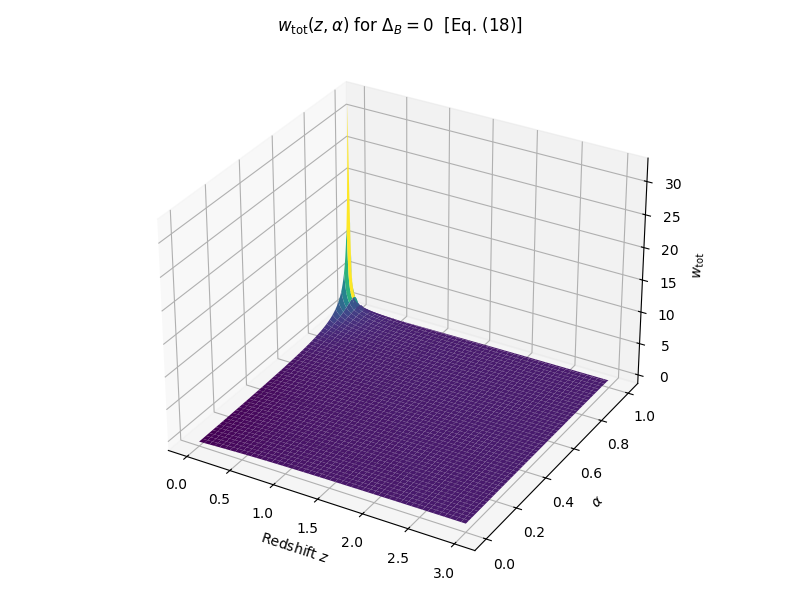}
        \caption{$\Delta_B = 0$}
    \end{subfigure}
    \hfill
    \begin{subfigure}[b]{0.32\textwidth}
        \centering
        \includegraphics[width=\textwidth]{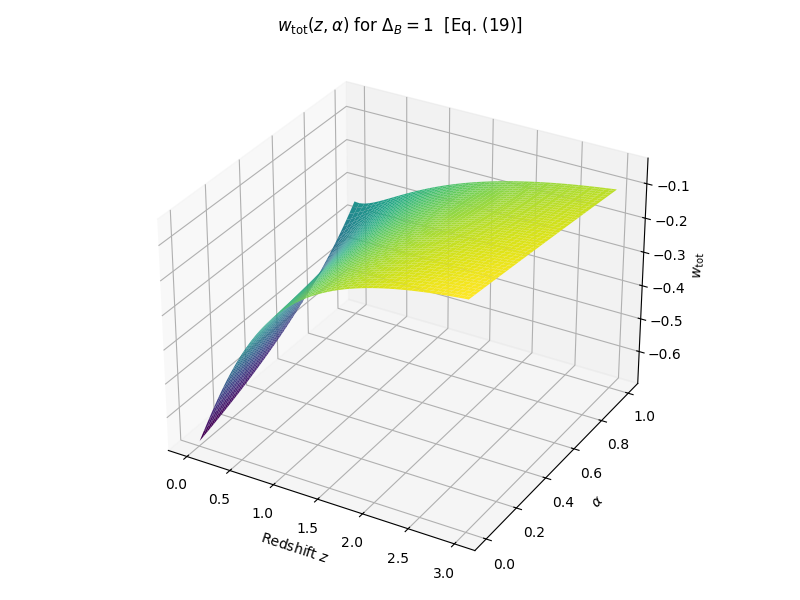}
        \caption{$\Delta_B = 1$}
    \end{subfigure}
    \hfill
    \begin{subfigure}[b]{0.32\textwidth}
        \centering
        \includegraphics[width=\textwidth]{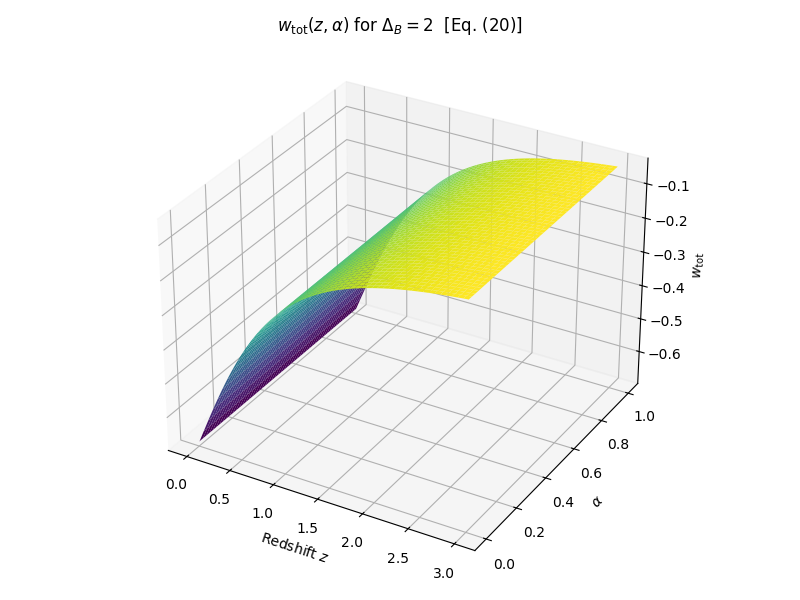}
        \caption{$\Delta_B = 2$}
    \end{subfigure}

    \caption{Three-dimensional evolution of the total equation-of-state parameter
    $w_{\mathrm{tot}}(z,\alpha)$ as a function of the redshift $z$ and parameter $\alpha$ for different values of the Barrow exponent
    $\Delta_B$. The plots correspond to Eqs.~\eqref{eq:E_delta0}, \eqref{eq:E_delta1}, and \eqref{eq:E_delta2}, respectively.
    }
   \label{fig:wtot_3D_alpha}
\end{figure}

Fig.~\ref{fig:wtot_3D_alpha} depicts the evolutionary pattern of the total EoS parameter $w_{\mathrm{tot}}(z,\alpha)$.  For the case $\Delta_B=0$, where Barrow deformation is absent, the model becomes a QCD ghost–like dark energy with a quadratic holographic correction. Furthermore, the behaviour of the surface indicates a sharp increase at low redshift for $\alpha \to 1$. This indicates that this region of the parameter $\alpha$ is not phenomenologically viable. Hence, in general we can say that in absence of deformation parameter, the cosmic dynamics is highly sensitive to the holographic coupling parameter $\alpha$. 
On the other hand, the Barrow-deformed scenarios $\Delta_B=1$ and $\Delta_B=2$ remain well-behaved over the entire parameter space, with $w_{\mathrm{tot}}<-1/3$ at late times and no phantom crossing, thereby confirming the physical viability and stability of the model.

\begin{figure}[H]
    \centering

    \begin{subfigure}[b]{0.32\textwidth}
        \centering
        \includegraphics[width=\textwidth]{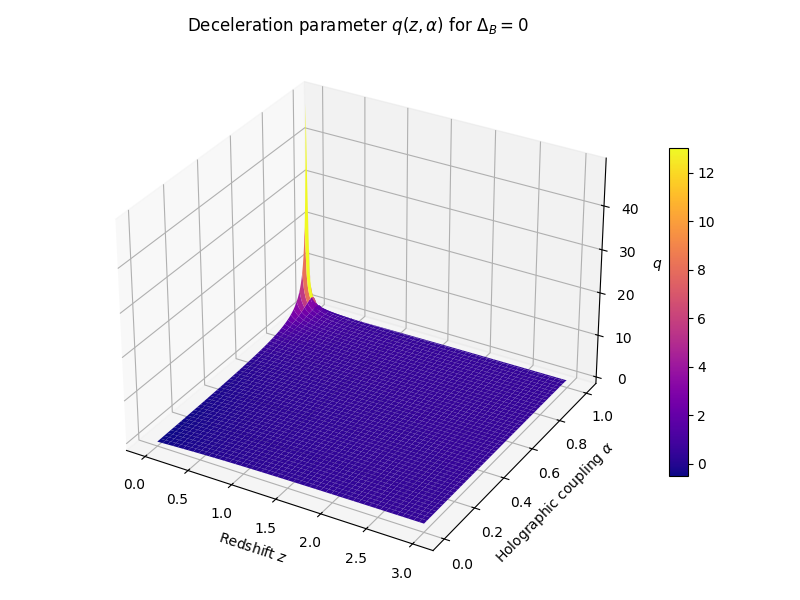}
        \caption{$\Delta_B = 0$}
    \end{subfigure}
    \hfill
    \begin{subfigure}[b]{0.32\textwidth}
        \centering
        \includegraphics[width=\textwidth]{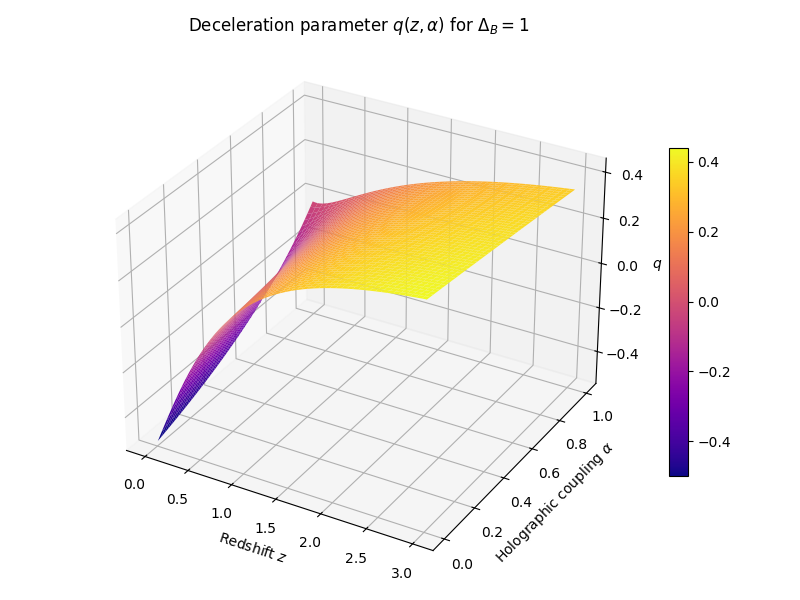}
        \caption{$\Delta_B = 1$}
    \end{subfigure}
    \hfill
    \begin{subfigure}[b]{0.32\textwidth}
        \centering
        \includegraphics[width=\textwidth]{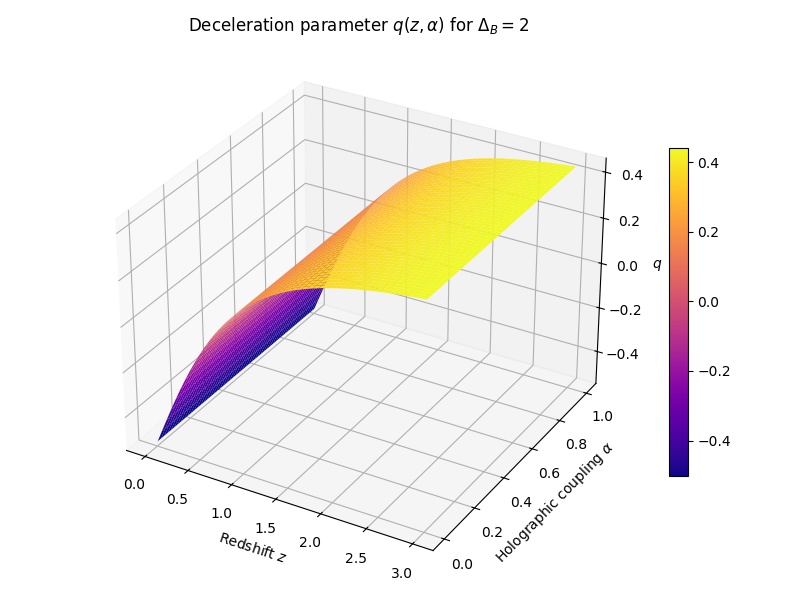}
        \caption{$\Delta_B = 2$}
    \end{subfigure}

    \caption{Three-dimensional evolution of the deceleration parameter
    $q(z,\alpha)$ as a function of the redshift $z$ and the holographic coupling
    parameter $\alpha$ for different values of the Barrow exponent $\Delta_B$,
    corresponding to Eqs.~\eqref{eq:E_delta0}, \eqref{eq:E_delta1} and the definition in Eq.~\eqref{q}.}

    \label{fig:q_3D_alpha}
\end{figure}

Fig. \ref{fig:q_3D_alpha} depicts the behaviour of the deceleration parameter with $(z,\alpha)$. In case of $\Delta_B=0$, we observe a sharp increase for high $\alpha$ and low redshift. A closer look following the colour bar shows that for the rest of the period the $q$ is flat and lies slightly below $0$. This indicates an unphysical behaviour. In contrast, if we look at the cases of $\Delta_B=1$ and $\Delta_B=2$, we observe weak variation with $\alpha$. However, the surface is smooth and monotonic is pattern. Furthermore, at high redshifts $q>0$ and ar low redshifts $q<0$. This indicates a clear transition from a decelerated to an accelerated expansion scenario. The weak dependence on $\alpha$ and the absence of any sudden changes in the Barrow-deformed cases imply the stabilizing role of entropy deformation parameter, resulting in a physically viable and observationally consistent description of the late-time cosmic acceleration.   

\section{Scalar-Field Reconstruction and Dynamical Interpretation of the BH–QCDGDE Model}

In this section, we will present a reconstruction scheme for the effective dark energy sector of the BH--QCDGDE model in terms of an equivalent scalar-field description by following the steps outlined below. We begin by assuming that the effective dark energy component can be described by a canonical scalar field minimally coupled to gravity. Based on the effective dark energy density $\rho_{\mathrm{DE}}$ and pressure $p_{\mathrm{DE}}$ obtained from the background cosmological dynamics elaborated in the previous section, we would now identify the scalar-field kinetic term and potential through the standard relations.  Assuming the dark energy component behaves as a canonical scalar field $\phi$, the kinetic term and potential are reconstructed as
\begin{align}
\dot{\phi}^2 &= \rho_{\mathrm{DE}}(1+w_{\mathrm{DE}}) \label{phidotsqr},\\
V(\phi) &= \frac{1}{2}\rho_{\mathrm{DE}}(1-w_{\mathrm{DE}}) \label{V},
\end{align}
allowing the model to be interpreted dynamically in terms of quintessence-like behavior.

Using the dark energy density of the BH--QCDGDE model, presented in Eq. \eqref{rhoDE} together with the equation-of-state parameter presented in Eq. \eqref{wDE} the scalar-field kinetic term takes the explicit form
\begin{equation}
\dot{\phi}^{2}
=
-3M_{\mathrm{Pl}}^{2}
\left[\alpha(2-\Delta_B)H^{1-\Delta_B}+\beta\right]
\frac{\dot{H}}{H}.
\label{phidot_explicit}
\end{equation}

Using the redshift-time relation $\dot{H}=-(1+z)H(z)\frac{dH(z)}{dz}$ in \eqref{phidot_explicit} we obtain

\begin{equation}
\dot{\phi}^{2}(z)
=
3M_{\mathrm{Pl}}^{2}
\left[\alpha(2-\Delta_B)H^{1-\Delta_B}(z)+\beta\right]
(1+z)\frac{dH(z)}{dz}.
\label{phidot_z}
\end{equation}

We know that $\frac{dH}{dz}>0$ is satisfied for an expanding universe. Also,  we have positive values of $\alpha$ and $\beta$. Hence, we automatically have the condition $\dot{\phi}^{2}>0$ to be satisfied. This indicates that the reconstructed scalar field exhibits quintessence-like behavior and avoids any phantom instability, as evident from \eqref{phidot_z}.

Let us now consider $\dot{\phi}=-(1+z)H(z)\frac{d\phi}{dz}$. Then on integration on \eqref{phidot_z} we obtain the scalar field evolving in the form 
\begin{equation}
\phi(z) = \sqrt{3}\,M_{\mathrm{Pl}}
\int_{0}^{z}\sqrt{\frac{\alpha(2-\Delta_B)H^{1-\Delta_B}(z')+\beta}
{(1+z')H^{2}(z')}\left(\frac{dH(z')}{dz'} \right)}\,dz'+C_0
\label{phi_z}
\end{equation}
where $C_0$ is constant of integration. The potential $V(\phi)$ is reconstructed using \eqref{V} and we get 
\begin{equation}
V(z)=3M_{\mathrm{Pl}}^{2}
\left(\alpha H^{2-\Delta_B}(z)+\beta H(z)\right)
-\frac{3M_{\mathrm{Pl}}^{2}}{2}
\left[\alpha(2-\Delta_B)H^{1-\Delta_B}(z)+\beta\right]
(1+z)\frac{dH(z)}{dz}.
\label{V_z}
\end{equation}

\begin{figure}[htbp]
    \centering

    \begin{subfigure}[b]{0.48\textwidth}
        \centering
        \includegraphics[width=\textwidth]{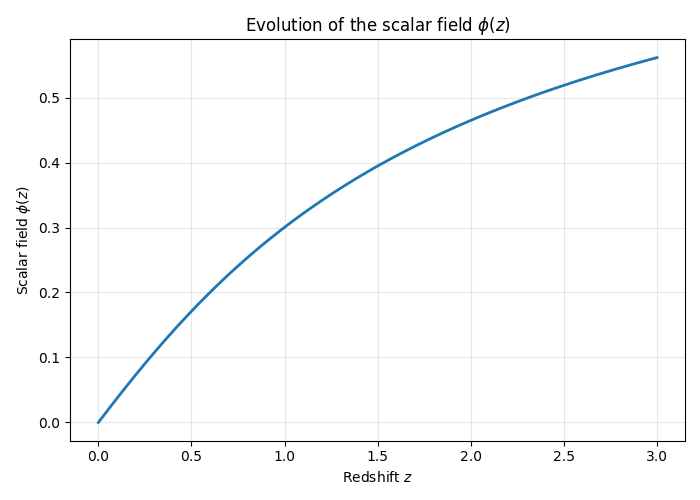}
        \caption{Evolution of the scalar field $\phi(z)$.}
    \end{subfigure}
    \hfill
    \begin{subfigure}[b]{0.48\textwidth}
        \centering
        \includegraphics[width=\textwidth]{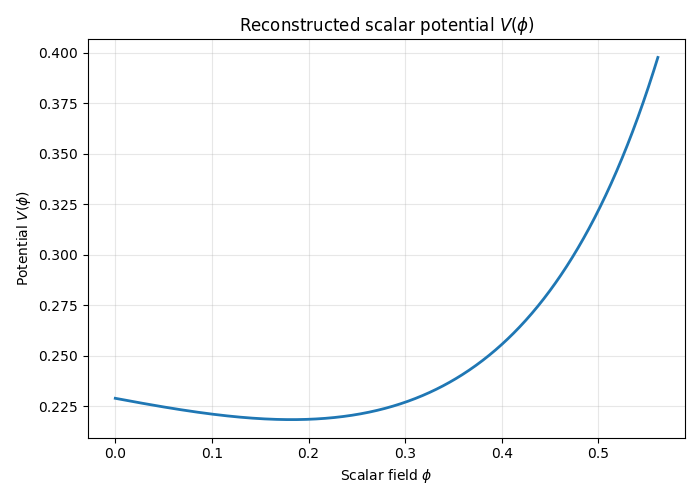}
        \caption{Reconstructed scalar potential $V(\phi)$.}
    \end{subfigure}

    \caption{Scalar-field reconstruction of the BH--QCDGDE model.
    The left panel shows the evolution of the scalar field $\phi$ as a
    function of the redshift $z$. The right panel displays the reconstructed parametric plot of scalar potential $V(\phi)$.}
        \label{fig:scalar_reconstruction}
\end{figure}

Fig.~\ref{fig:scalar_reconstruction} illustrates the scalar-field reconstruction of the BH--QCDGDE model, showing a smooth and monotonic evolution of the scalar field $\phi(z)$ with redshift together with the reconstructed potential $V(\phi)$. The panel (a) shows that $\phi(z)$ is showing a monotone increasing behaviour with $z$. The potential $V(\phi)$ exhibits a slowly decreasing profile at low redshifts. After certain redshift $V(\phi)$ starts increasing with $z$. Overall, from Fig.~\ref{fig:scalar_reconstruction} we can say that the scalar-field description remains free from phantom instabilities and provides a consistent dynamical interpretation of cosmic acceleration in the BH--QCDGDE framework.

\section{Analytical Power-Law Solutions and Scalar-Field Dynamics in the BH–QCDGDE Model}

In ths section, we consider a particular case of power-law scale factor. This is being added to gain further analytical insight into the cosmological behaviour of the BH--QCDGDE model. In this context, we consider a particular solution corresponding to a power-law expansion of the scale factor. It is well establshed in the literature that power-law cosmologies provide a useful approximation to various evolutionary phases of the universe and allow transparent interpretation of the underlying dynamics.

For this purpose, we assume the scale factor in the form
\begin{equation}
a(t) = a_{0}\, t^{n}, \qquad n>0 ,
\label{plaw}
\end{equation}
where $n>1$ is a dimensionless constant. The corresponding Hubble parameter and its time derivative are given by
\begin{equation}
H(t) = \frac{\dot a}{a} = \frac{n}{t}, 
\qquad 
\dot H = -\frac{n}{t^{2}} .
\end{equation}

Substituting $H=\frac{n}{t}$ into the BH--QCDGDE density \eqref{rhoDE}, we obtain
\begin{equation}
\rho_{\rm DE}(t) = 3M_{\rm Pl}^{2}\left[
\alpha \left(\frac{n}{t}\right)^{2-\Delta_{B}}
+ \beta \left(\frac{n}{t}\right)
\right].
\label{rho_pl}
\end{equation}
This expression explicitly shows how the Barrow entropy deformation and the QCD ghost contribution scale with cosmic time in the case of a power-law form of scale factor.

Using the conservation equation together with Eq.~\eqref{wDE}, the effective dark-energy equation-of-state parameter for the power-law case becomes
\begin{equation}
w_{\rm DE}
= -1 + \frac{1}{n}\left(
\frac{\alpha (2-\Delta_{B}) n^{1-\Delta_{B}} t^{\Delta_{B}-1} + \beta}
{\alpha n^{2-\Delta_{B}} t^{\Delta_{B}-2} + \beta n t^{-1}}\right) .
\label{w_pl}
\end{equation}
For $n>1$, corresponding to accelerated expansion, the above expression remains greater than $-1$ as $\frac{1}{n}\left(
\frac{\alpha (2-\Delta_{B}) n^{1-\Delta_{B}} t^{\Delta_{B}-1} + \beta}
{\alpha n^{2-\Delta_{B}} t^{\Delta_{B}-2} + \beta n t^{-1}}\right)>0$, indicating a quintessence-like behaviour of the effective dark energy sector.

The scalar field and potentials automatically come out to be as follows:

\begin{equation}
\dot{\phi}^{\,2}(t)
= 3 M_{\rm Pl}^{2}
\left[
\alpha (2-\Delta_{B})\,\frac{n^{2-\Delta_{B}}}{t^{3-\Delta_{B}}}
+ \beta\,\frac{n}{t^{2}}
\right].
\label{phidot2_pl}
\end{equation}

\begin{equation}
\phi(t)
= \phi_{0}
+ \sqrt{3}\,M_{\rm Pl}
\int
\sqrt{
\alpha (2-\Delta_{B})\,\frac{n^{2-\Delta_{B}}}{t^{3-\Delta_{B}}}
+ \beta\,\frac{n}{t^{2}}
}\,dt .
\label{phi_pl}
\end{equation}

\begin{equation}
V(t)
= \frac{3 M_{\rm Pl}^{2}}{2}
\left[
\alpha \Delta_{B}\,\frac{n^{2-\Delta_{B}}}{t^{2-\Delta_{B}}}
+ \beta\,\frac{n}{t}
\right].
\label{V_pl}
\end{equation}

\begin{figure}[htbp]
\centering
\begin{subfigure}{0.48\textwidth}
    \centering
    \includegraphics[width=\linewidth]{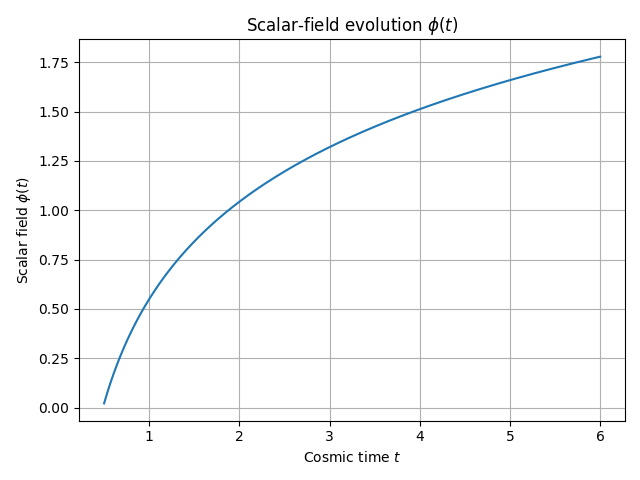}
    \caption{Evolution of the scalar field $\phi(t)$}
    \label{fig:phi_t}
\end{subfigure}
\hfill
\begin{subfigure}{0.48\textwidth}
    \centering
    \includegraphics[width=\linewidth]{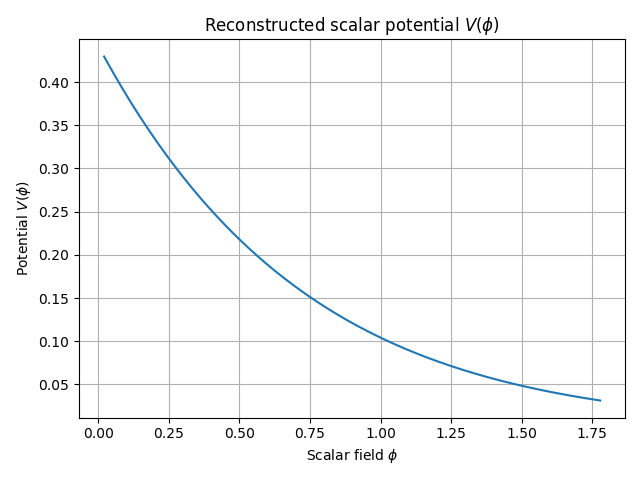}
    \caption{Reconstructed scalar potential $V(\phi)$}
    \label{fig:V_phi}
\end{subfigure}
\caption{Scalar-field reconstruction for the power-law scale factor $a(t)\propto t^{n}$. 
The left panel shows the monotonic evolution of the scalar field $\phi(t)$ obtained from Eq.~\eqref{phi_pl}, 
while the right panel depicts the corresponding reconstructed potential $V(\phi)$ from Eq.~\eqref{V_pl}. 
The plots are generated for the representative parameter values 
$\alpha=0.02$, $\beta=0.08$, $\Delta_{B}=0.4$, $n=1.5$, with the reduced Planck mass set to $M_{\rm Pl}=1$.}
\label{fig:scalar_reconstruction}
\end{figure}

Fig.~\ref{fig:scalar_reconstruction} pictorially presents the scalar-field reconstruction of the BH--QCDGDE model for a power-law scale factor $a(t)\propto t^{n}$. In the left panel of Fig.~\ref{fig:scalar_reconstruction} we observe that the scalar field $\phi(t)$ evolves monotonically with $t$, indicating a rolling scalar field . We observe a gradual flattening of the $\phi(t)$ curve at late times. This can be interpreted as decreasing influence of Barrow deformation parameter and the increasing influence of the linear QCD ghost contribution with the evolution of the universe. In the right panelwe plot the reconstructed scalar potential $V(\phi)$, which is a smooth and monotonically decreasing function of the field $\phi$.  

\section{Scale-Factor Reconstruction in the BH–QCDGDE Framework}
In this section, our focus is the late-time cosmological evolution implied by the BH--QCDGDE model. We intend to do it by deriving an explicit form of the scale factor. Employing the linearized background equations for a small Barrow deformation, we derive an analytical expression for the Hubble parameter and the corresponding scale factor. Let us begin with the following expansion
\begin{equation}
\begin{aligned}
H^{2-\Delta_B}
=\, & H^2
\Bigg[
1
-\Delta_B \ln H
+\frac{\Delta_B^{2}}{2!}\left(\ln H\right)^{2}
-\frac{\Delta_B^{3}}{3!}\left(\ln H\right)^{3}  \\
& \qquad
+\frac{\Delta_B^{4}}{4!}\left(\ln H\right)^{4}
+\mathcal{O}\!\left(\Delta_B^{5}\right)
\Bigg].
\end{aligned}
\label{expansion}
\end{equation}

Considering $\Delta_B \ll 1$ in \eqref{expansion} i.e. a small Barrow deformation parameter and therefore retaining terms up to linear order in $\Delta_B$, we obtain
\begin{equation}
H^{2-\Delta_B}
\simeq H^2 \left( 1 - \Delta_B \ln H \right),
\qquad (\Delta_B \ll 1).
\label{linexp}
\end{equation}

Substituting Eq.~\eqref{linexp} into the first Friedmann equation, we can write
\begin{equation}
3 M_{\rm Pl}^2 H^2
= \rho_m + \rho_r
+ 3 M_{\rm Pl}^2 \left( \alpha H^{2-\Delta_B} + \beta H \right),
\end{equation}
This leads to the linearized background equation
\begin{equation}
(1-\alpha) H^2 - \beta H
- \frac{\rho_m + \rho_r}{3 M_{\rm Pl}^2}
+ \alpha \Delta_B H^2 \ln H = 0,
\qquad (\Delta_B \ll 1).
\label{linfried}
\end{equation}

We introduce the dimensionless Hubble parameter
\begin{equation}
E(z) \equiv \frac{H(z)}{H_0}.
\label{Ez0}
\end{equation}
The present-day density parameters
\begin{equation}
\Omega_{m0} \equiv \frac{\rho_{m0}}{3 M_{\rm Pl}^2 H_0^2},
\qquad
\Omega_{r0} \equiv \frac{\rho_{r0}}{3 M_{\rm Pl}^2 H_0^2}.
\end{equation}
Using $\rho_m=\rho_{m0}(1+z)^3$ and $\rho_r=\rho_{r0}(1+z)^4$, Eq.~\eqref{linfried}
can be written in the following form
\begin{equation}
(1-\alpha) E^2(z) - \beta E(z)
- \Omega_{m0}(1+z)^3 - \Omega_{r0}(1+z)^4
+ \alpha \Delta_B E^2(z) \ln E(z) = 0.
\label{linEz}
\end{equation}

As we are assuming that the Barrow deformation parameter satisfies $\Delta_B \ll 1$, we apply Taylor series expansion around $\Delta_B=0$ and $E(z)$ can be expanded as $E(z,\Delta_B) =E(z,0)+\left.\frac{\partial E}{\partial \Delta_B}\right|_{\Delta_B=0}\Delta_B+
\mathcal{O}\!\left(\Delta_B^{2}\right).$ We define zeroth-order solution as $E_0(z)\equiv E(z)\big|_{\Delta_B=0}$ and and the
leading Barrow-induced correction as $E_1(z)\equiv \left.\partial E/\partial \Delta_B\right|_{\Delta_B=0}$. Now, to first order in $\Delta_B$, we write
\begin{equation}
E(z) = E_0(z) + \Delta_B E_1(z),
\end{equation}
Accordingly, we have the linearized Friedmann equation,
\begin{equation}
(1-\alpha)E^2-\beta E
-\Omega_{m0}(1+z)^3-\Omega_{r0}(1+z)^4
+\alpha\Delta_B E^2\ln E=0,
\end{equation}
and keeping terms up to first order in $\Delta_B$, we can obtain
\begin{equation}
(1-\alpha)\!\left(E_0^2+2\Delta_B E_0E_1\right)
-\beta\!\left(E_0+\Delta_B E_1\right)
-\Omega_{m0}(1+z)^3-\Omega_{r0}(1+z)^4
+\alpha\Delta_B E_0^2\ln E_0=0 .
\label{eq:lin}
\end{equation}
It may be noted that to obtain \eqref{eq:lin} we have used $\ln\!\left(E_0+\Delta_B E_1\right)
=\ln\!\left[E_0\left(1+\Delta_B\frac{E_1}{E_0}\right)\right]=\ln E_0 + \ln\!\left(1+\Delta_B\frac{E_1}{E_0}\right)$.
Finally, the dimensionless Hubble parameter up to linear order in $\Delta_B$ comes out to be
\begin{equation}
E(z)=
E_0(z)
-\Delta_B
\frac{\alpha E_0^2(z)\ln E_0(z)}
{2(1-\alpha)E_0(z)-\beta},
\qquad (\Delta_B\ll1).
\label{Elinsol}
\end{equation}
Hence, the Hubble parameter up to linear order in $\Delta_B$ is
\begin{equation}
H(z)=
H_0 E_0(z)
-
\Delta_B H_0
\frac{\alpha E_0^2(z)\ln E_0(z)}
{2(1-\alpha)E_0(z)-\beta},
\qquad (\Delta_B\ll1),
\label{Hlinsol}
\end{equation}
where $E_0(z)=\frac{\beta+\sqrt{\beta^2+4(1-\alpha) \left[\Omega_{m0}(1+z)^3+\Omega_{r0}(1+z)^4\right]}} {2(1-\alpha)}$.
Finally, we have the solution for scale factor
\begin{equation}
a(t)=a_0
\exp\!\left[\left(\frac{\beta}{1-\alpha}-\Delta_B
\frac{\alpha \beta}{2(1-\alpha)^2}\ln\!\left(\frac{\beta}{1-\alpha}\right)
\right)(t-t_0)\right].
\label{asol}
\end{equation}
Using Eqs.~\eqref{Hlinsol} and \eqref{asol} in Eq.~\eqref{w} we obtain the reconstructed EoS parameter
\begin{equation}
w_{\rm tot}(z)
=-1+\frac{2(1+z)}{3E(z)} \left[\frac{dE_0(z)}{dz}
-\Delta_B \frac{d}{dz}
\left(\frac{\alpha E_0^{2}(z)\ln E_0(z)}
{2(1-\alpha)E_0(z)-\beta}
\right)
\right].
\label{50}
\end{equation}

In the present-epoch limit i.e. $z\to0$, Eq.~\eqref{50} reduces to
\[
w_{\rm tot}(0)= -1 + \frac{2}{3}\left.\frac{dE}{dz}\right|_{z=0},
\]
since $E(0)=1$. As the Hubble rate varies slowly at late times, we have
$0<\left.\frac{dE}{dz}\right|_{0}\ll1$, implying $-1<w_{\rm tot}(0)<-1/3$ and hence consistent with late-time acceleration of the universe.

Using the reconstructed Hubble parameter obtained in Eq.~\eqref{Hlinsol}, and substituting it into the BH--QCDGDE density definition \eqref{rhoDE}, we derive the reconstructed form of the BH--QCDGDE density for the scale factor obtained in \eqref{asol} as follows:
\begin{equation}
\rho_{\rm DE}(H)=3M_{\rm Pl}^{2}\left[\alpha H^{2-\Delta_B}+\beta H\right].
\label{rho1}
\end{equation}
We can rewrite \eqref{rho1} in terms of redshift $z$ as
\begin{equation}
\begin{aligned}
\rho_{\rm DE}(z)
&=3M_{\rm Pl}^{2}\Bigg\{
\alpha\!\left[
H_0E_0(z)
-\Delta_BH_0
\frac{\alpha E_0^{2}(z)\ln E_0(z)}
{2(1-\alpha)E_0(z)-\beta}
\right]^{2-\Delta_B}
\\
&\hspace{1.8cm}
+\beta\!\left[
H_0E_0(z)
-\Delta_BH_0
\frac{\alpha E_0^{2}(z)\ln E_0(z)}
{2(1-\alpha)E_0(z)-\beta}
\right]
\Bigg\},
\end{aligned}
\label{rho_exact}
\end{equation}
where
\begin{equation}
E_0(z)=
\frac{\beta+\sqrt{\beta^2+4(1-\alpha)\left[\Omega_{m0}(1+z)^3+\Omega_{r0}(1+z)^4\right]}}
{2(1-\alpha)} .
\end{equation}

For a small Barrow deformation parameter, after retaining terms up to first order in
\(\Delta_B\), we obtain the reconstructed dark energy density in a simplified form as:
\begin{equation}
\rho_{\rm DE}(z)\simeq
3M_{\rm Pl}^{2}H_0^{2}
\left[
\alpha E_0^{2}(z)+\beta E_0(z)
-\Delta_B\,\alpha E_0^{2}(z)\ln E_0(z)
\right].
\label{rho_linear}
\end{equation}

From the form of the dark energy density reconstructed in Eq.~\eqref{rho_linear} we can see that the reconstructed dark energy density consists of three distinct contributions. These contributions are, namely, a holographic-like term proportional to \(E_0^{2}(z)\), a QCD ghost term linear in \(E_0(z)\), and a logarithmic correction induced by the Barrow entropy deformation. At later epochs of the universe, i.e. (\(z\to0\), \(E_0\to1\)), the logarithmic term present in the expression Eq.~\eqref{rho_linear}  becomes less effective. This implies that \(\rho_{\ rmDE}\) can asymptotically approach a cosmological-constant-like behaviour. As a consequence, in the BH--QCDGDE we can obtain a stable late-time accelerated expansion in alignment with the reconstructed Hubble evolution.

Using the reconstructed total equation-of-state parameter obtained in Eq.~\eqref{50}, together with the reconstructed dark-energy density (Eq.~\eqref{rho_linear}), the scalar-field kinetic term becomes the following:
\begin{equation}
\begin{array}{rcl}
\dot{\phi}^{2}(z)
&=&
2M_{\rm Pl}^{2}H_0^{2}(1+z)
\left[
\alpha E_0^{2}(z)+\beta E_0(z)
-\Delta_B\,\alpha E_0^{2}(z)\ln E_0(z)
\right]
\\[6pt]
&&\times
\displaystyle
\frac{1}{E(z)}
\left[
\frac{dE_0(z)}{dz}
-\Delta_B
\frac{d}{dz}
\left(
\frac{\alpha E_0^{2}(z)\ln E_0(z)}
{2(1-\alpha)E_0(z)-\beta}
\right)
\right].
\end{array}
\label{phi_dot}
\end{equation}

Using $\dot{\phi}=-(1+z)H(z)\,d\phi/dz$, the scalar field evolves as
\begin{equation}
\begin{array}{rcl}
\phi(z)
&=&
\sqrt{2}\,M_{\rm Pl}
\displaystyle
\int
\Bigg\{
\frac{
\alpha E_0^{2}(z)+\beta E_0(z)
-\Delta_B\,\alpha E_0^{2}(z)\ln E_0(z)
}{E(z)}
\\[6pt]
&&\hspace{1.0cm}\times
\left[
\frac{dE_0(z)}{dz}
-\Delta_B
\frac{d}{dz}
\left(
\frac{\alpha E_0^{2}(z)\ln E_0(z)}
{2(1-\alpha)E_0(z)-\beta}
\right)
\right]
\Bigg\}^{1/2}
\,dz
+ C .
\end{array}
\label{phi_z}
\end{equation}

where $C$ is an integration constant. In Eq.~\eqref{V} we use \eqref{phi_dot} and \eqref{50} and obtain the reconstructed scalar potential as
\begin{equation}
\begin{array}{rcl}
V(z)
&=&
\displaystyle
\frac{3}{2}M_{\rm Pl}^{2}H_0^{2}
\left[
\alpha E_0^{2}(z)+\beta E_0(z)
-\Delta_B\,\alpha E_0^{2}(z)\ln E_0(z)
\right]
\\[6pt]
&&\times
\left[
2-
\frac{2(1+z)}{3E(z)}
\left(
\frac{dE_0(z)}{dz}
-\Delta_B
\frac{d}{dz}
\left(
\frac{\alpha E_0^{2}(z)\ln E_0(z)}
{2(1-\alpha)E_0(z)-\beta}
\right)
\right)
\right].
\end{array}
\label{Vz}
\end{equation}

From the forms of Eq.~\eqref{phi_z} and \eqref{Vz} we can interpret that the reconstructed scalar field is expected to have a monotone pattern with redshift. A logarithmic corrections to both the kinetic term and the potential is induced by $\Delta_B$  i.e. the Barrow deformation parameter. This reflects the quantum-gravitational effect in the effective scalar description. At later phases of the universe i.e. ($z\to0$, $E_0\to1$), the potential approaches a slowly varying form that implies an approach toward a cosmological-constant-like regime.

\begin{figure}[htbp]
\centering

\begin{subfigure}[b]{0.48\linewidth}
\centering
\includegraphics[width=\linewidth]{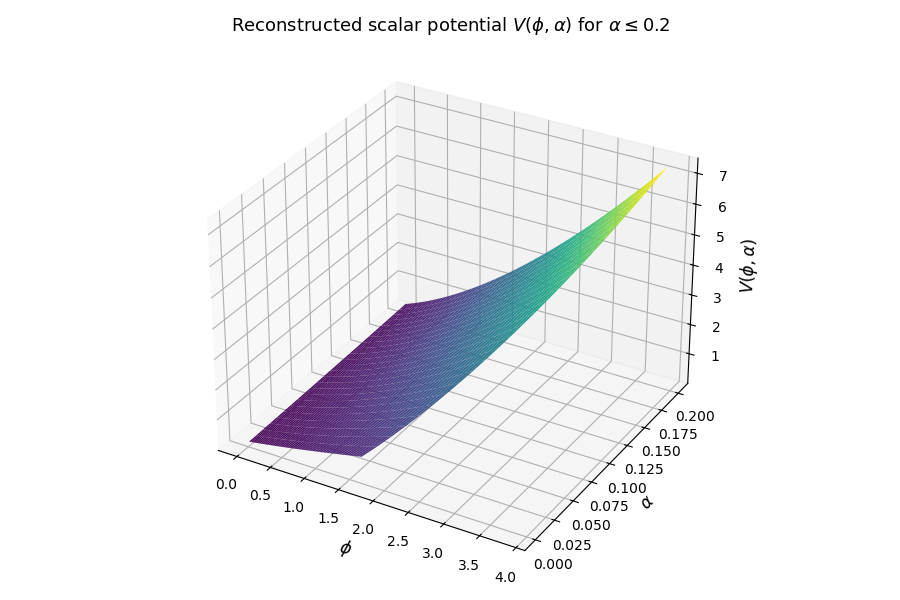}
\caption{$V(\phi,\alpha)$ for $\alpha \leq 0.2$.}
\label{fig:V_phi_alpha}
\end{subfigure}
\hfill
\begin{subfigure}[b]{0.48\linewidth}
\centering
\includegraphics[width=\linewidth]{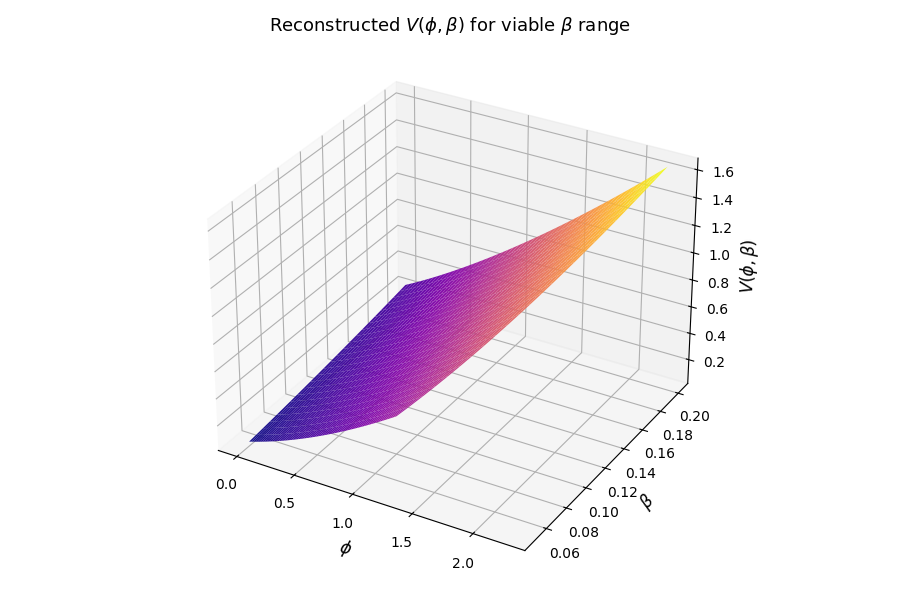}
\caption{$V(\phi,\beta)$ for $0.05 \leq \beta \leq 0.2$.}
\label{fig:V_phi_beta}
\end{subfigure}

\vspace{0.8em}

\begin{subfigure}[b]{0.65\linewidth}
\centering
\includegraphics[width=\linewidth]{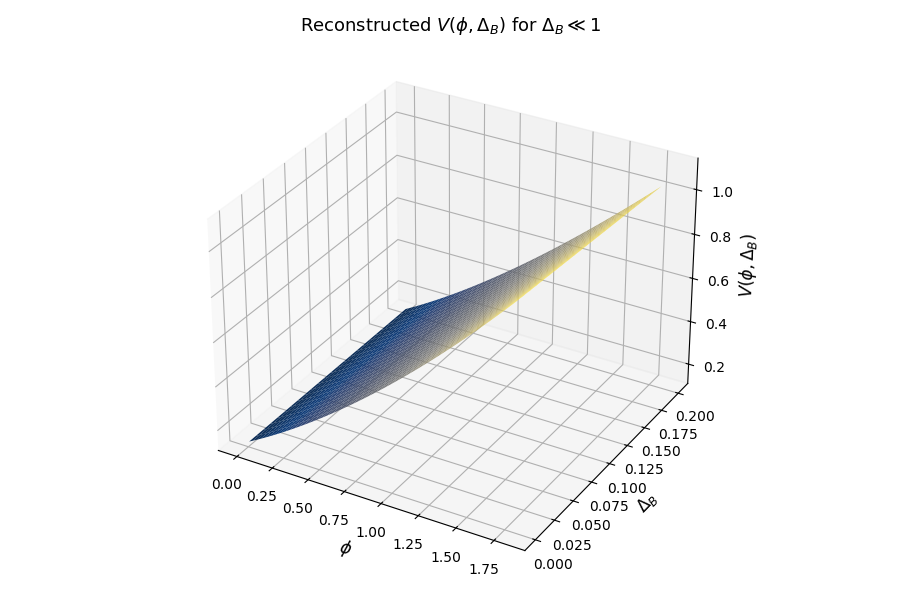}
\caption{$V(\phi,\Delta_B)$ for the small-deformation regime $\Delta_B \ll 1$.}
\label{fig:V_phi_DeltaB}
\end{subfigure}

\caption{
Reconstructed scalar potential $V(\phi)$ as a function of the scalar field
$\phi$ and the model parameters.
\textbf{Top left}: variation with the holographic parameter $\alpha$ in the
observationally viable range $\alpha \leq 0.2$, for fixed
$\Omega_{m0}=0.3$, $\Omega_{r0}=8\times10^{-5}$, $\beta=0.12$, and
$\Delta_B=0.1$.
\textbf{Top right}: variation with the QCD ghost parameter $\beta$ in the range
$0.05 \leq \beta \leq 0.2$, keeping
$\Omega_{m0}=0.3$, $\Omega_{r0}=8\times10^{-5}$, $\alpha=0.01$, and
$\Delta_B=0.1$ fixed.
\textbf{Bottom}: dependence on the Barrow deformation parameter $\Delta_B$ in
the perturbative regime $\Delta_B \ll 1$, with
$\Omega_{m0}=0.3$, $\Omega_{r0}=8\times10^{-5}$, $\alpha=0.01$, and
$\beta=0.12$.
In all panels, natural units $H_0=1$ and $M_{\rm Pl}=1$ are adopted.
}
\label{fig:V_phi_reconstruction}
\end{figure}

In Fig.~\ref{fig:V_phi_reconstruction} we have presented the reconstructed scalar potential $V(\phi)$ based on \eqref{phi_z} and \eqref{Vz} for the BH--QCDGDE model. The 3D surfaces corresponding to variations of $\alpha$ and $\beta$ are demonstrating a smooth and monotonic dependence on the scalar field. Here, $\alpha$ is the holographic coupling and $\beta$ is the QCD ghost parameter. It is further observed in Figure~7 that $V(\phi)$ steepens mildly with $\alpha$ and $\beta$. This indicates enhanced dark-energy dominance and the potential $V(\phi)$ is free from any sudden changes within the observationally viable ranges of the parameters. On the other hand, the third panel of Fig.~\ref{fig:V_phi_reconstruction} reflects that the sensitivity of the scalar potential $V(\phi)$ appears to be relatively weak for the Barrow deformation parameter $\Delta_B \ll 1$.  This confirms that quantum-gravitational entropy corrections introduce only perturbative modifications, preserving a stable quintessence-like late-time dynamics.

\section{Generalized Second Law of Thermodynamics}

In this section, we examine the validity of the generalized second law of thermodynamics (GSLT) within the framework of the BH--QCDGDE model. The GSLT states that the total entropy of the Universe, defined as the sum of the entropy of the apparent horizon and that of the matter--energy content enclosed within it, must be a non-decreasing function of cosmic time \cite{KM2024,Mazumder2010,Kumar2021,Mathew2014,Setare2007}
,
\begin{equation}
\dot{S}_{\rm tot}=\dot{S}_{h}+\dot{S}_{\rm in}\geq 0 .
\end{equation}

\subsection{Apparent horizon entropy}

For a spatially flat FLRW universe, the radius of the apparent horizon is given by \cite{JamilSaridakisSetare2010}
\begin{equation}
R_{A}=\frac{1}{H},
\end{equation}
with the associated temperature \cite{CaiKim2005, CaiCaoHu2009,HaldarBhattacharjeeChakraborty2017}
\begin{equation}
T_{h}=\frac{1}{2\pi R_{A}}=\frac{H}{2\pi}.
\end{equation}
The thermodynamics of the apparent horizon and the validity of the generalized
second law have been extensively investigated in various cosmological settings
\cite{AbreuNeto2022,CaiCao2007}. In the presence of Barrow entropy, the horizon entropy takes the form
\begin{equation}
S_{h}=\left(\frac{A}{A_{0}}\right)^{\frac{1+\Delta_{B}}{2}}
= \left(\frac{4\pi}{H^{2}}\right)^{\frac{1+\Delta_{B}}{2}},
\label{sh}
\end{equation}
where $A=4\pi R_{A}^{2}$ and natural units are assumed. Differentiating with respect to cosmic time, we obtain
\begin{equation}
\dot{S}_{h}
=-(1+\Delta_{B})(4\pi)^{\frac{1+\Delta_{B}}{2}}
H^{-(3+\Delta_{B})}\dot{H}.
\label{Sh_dot}
\end{equation}
Since $\dot{H}<0$ for an expanding universe, it follows immediately that $\dot{S}_{h}>0$.

\subsection{Entropy of the cosmic fluid}

The entropy variation of the matter--energy content inside the apparent horizon is governed by the Gibbs equation \cite{TianBooth2015},
\begin{equation}
T_{h} dS_{\rm in}=d(\rho_{\rm tot} V)+p_{\rm tot}\, dV,
\label{Th}
\end{equation}
where $V=\frac{4\pi}{3}R_{A}^{3}=\frac{4\pi}{3H^{3}}$ is the volume enclosed by the horizon. Using the continuity equation,
\begin{equation}
\dot{\rho}_{\rm tot}+3H(\rho_{\rm tot}+p_{\rm tot})=0,
\end{equation}
one finds \cite{Sheykhi2010}
\begin{equation}
\dot{S}_{\rm in}
=\frac{4\pi}{T_{h}H^{2}}(\rho_{\rm tot}+p_{\rm tot})\dot{H}.
\end{equation}
Employing the Friedmann equation,
\begin{equation}
\dot{H}=-\frac{1}{2M_{\rm Pl}^{2}}(\rho_{\rm tot}+p_{\rm tot}),
\end{equation}
the above expression reduces to
\begin{equation}
\dot{S}_{\rm in}
=-\frac{8\pi^{2}}{H^{3}}(\rho_{\rm tot}+p_{\rm tot}).
\label{S_indot}
\end{equation}

In accordance with the Friedmann equation, the total energy density of the Universe is
\begin{equation}
\rho_{\rm tot}
=
\rho_m+\rho_r+\rho_{\rm DE}.
\label{rhotot_def}
\end{equation}
For the BH--QCDGDE model, the dark-energy density is explicitly given by Eq.~\eqref{rhoDE}. Thus, the total energy density in terms of the Hubble parameter can be written as
\begin{equation}
\rho_{\rm tot}(H)=\rho_m+\rho_r+3M_{\rm Pl}^{2}\left(\alpha H^{2-\Delta_B}+\beta H\right).
\label{rhotot_H}
\end{equation}

Using the standard redshift evolution of matter and radiation,
\begin{equation}
\rho_m=\rho_{m0}(1+z)^3,
\qquad
\rho_r=\rho_{r0}(1+z)^4,
\label{rhomr_z}
\end{equation}
together with
\begin{equation}
\rho_{m0}=3M_{\rm Pl}^{2}H_0^{2}\Omega_{m0},
\qquad
\rho_{r0}=3M_{\rm Pl}^{2}H_0^{2}\Omega_{r0},
\label{rho0_defs}
\end{equation}
and considering the dimensionless Hubble parameter $E(z)$ as defined in Eq.~\eqref{Ez}. Then the total energy density can be recast into the redshift-dependent form
\begin{equation}
\rho_{\rm tot}(z)
=3M_{\rm Pl}^{2}H_0^{2}\left[\Omega_{m0}(1+z)^3+
\Omega_{r0}(1+z)^4 + \alpha E^{2-\Delta_B}(z) +
\beta E(z) \right].
\label{rhotot_z}
\end{equation}

This form of $\rho_{\rm tot}(z)$ is particularly convenient for the thermodynamic analysis, as it allows the generalized second law of thermodynamics to be examined directly in terms of the redshift evolution of the Hubble parameter. The total pressure is given by
$p_{\rm tot}=p_m+p_r+p_{\rm DE}
=\frac{1}{3}\rho_r+p_{\rm DE}$ (as $p_r=\frac{1}{3}\rho_r$), with $p_{\rm DE}=w_{\rm DE}\rho_{\rm DE}$ and $\rho_{\rm DE}$ defined in Eq.~\eqref{rhoDE}. Hence, finally

\begin{equation}
p_{\rm tot}(z)
=
M_{\rm Pl}^{2} H_{0}^{2}\,\Omega_{r0}(1+z)^{4}
+
3 M_{\rm Pl}^{2} H_{0}^{2}\,
w_{\rm DE}(z)
\left[
\alpha E^{2-\Delta_{B}}(z)+\beta E(z)
\right]
\label{p_tot}
\end{equation}

Now, using Eqs. \eqref{p_tot} and \eqref{rhotot_z} in \eqref{S_indot} we obtain

\begin{equation}
\dot S_{\rm in}
=
-\frac{24\pi M_{\rm Pl}^{2} H_0^{2}}{H^{3}}
\Bigg[\Omega_{m0}(1+z)^3
+\frac{4}{3}\Omega_{r0}(1+z)^4
+\big(1+w_{\rm DE}(z)\big)
\Big(\alpha E^{\,2-\Delta_B}(z)
+\beta E(z)\Big)\Bigg].
\label{Sin_dot}
\end{equation}

Using Eq.~\eqref{wDEEz} in \eqref{Sin_dot} we rewrite \eqref{Sin_dot} purely in terms of $E(z)$ as follows:

\begin{equation}
\dot S_{\rm in}
=-\frac{24\pi M_{\rm Pl}^{2} H_0^{2}}{H^{3}}
\Bigg[\Omega_{m0}(1+z)^3 +\frac{4}{3}\Omega_{r0}(1+z)^4
+(1+z) \Big(\alpha(2-\Delta_B)E^{1-\Delta_B}(z)+\beta
\Big) \frac{dE(z)}{dz}\Bigg].
\label{Sin_dot_E}
\end{equation}

Adding \eqref{Sh_dot} and \eqref{Sin_dot_E} to obtain the time derivative of total entropy $\dot{S}_{tot}$ in terms of $E(z)$ as
\begin{equation}
\begin{aligned}
\dot S_{\rm tot}
&=
(1+\Delta_B)(4\pi)^{\frac{1+\Delta_B}{2}}
H_0^{-(1+\Delta_B)}
(1+z)\,
E^{-(2+\Delta_B)}(z)
\frac{dE(z)}{dz}
\\[6pt]
&\quad
-\frac{24\pi M_{\rm Pl}^{2}}{H_0\,E^{3}(z)}
\Bigg[
\Omega_{m0}(1+z)^3
+\frac{4}{3}\Omega_{r0}(1+z)^4
\\[4pt]
&\qquad
+(1+z)
\Big(
\alpha(2-\Delta_B)E^{1-\Delta_B}(z)+\beta
\Big)
\frac{dE(z)}{dz}
\Bigg].
\end{aligned}
\label{Stot_dot_aligned}
\end{equation}

\begin{figure}[htbp]
\centering

\begin{subfigure}[b]{0.48\linewidth}
\centering
\includegraphics[width=\linewidth]{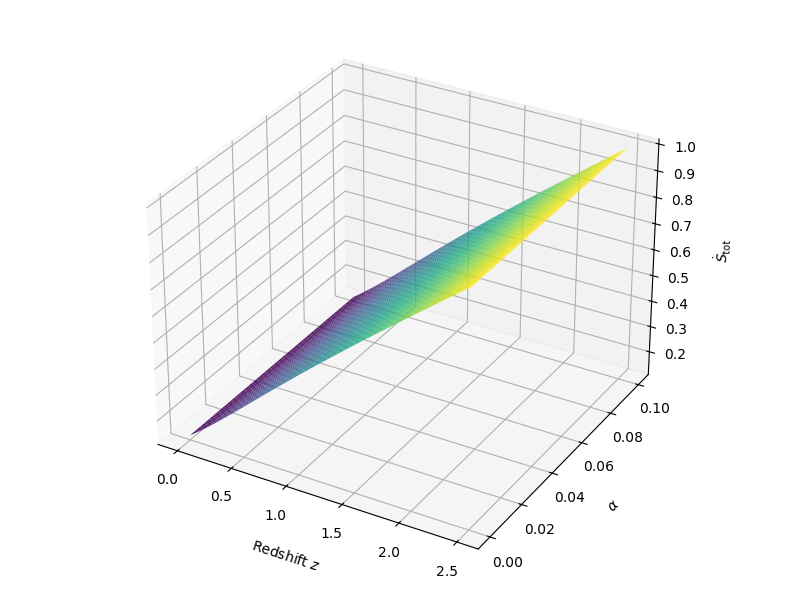}
\caption{$\dot S_{\rm tot}(z,\alpha)$}
\label{fig:Sdot_total_alpha}
\end{subfigure}
\hfill
\begin{subfigure}[b]{0.48\linewidth}
\centering
\includegraphics[width=\linewidth]{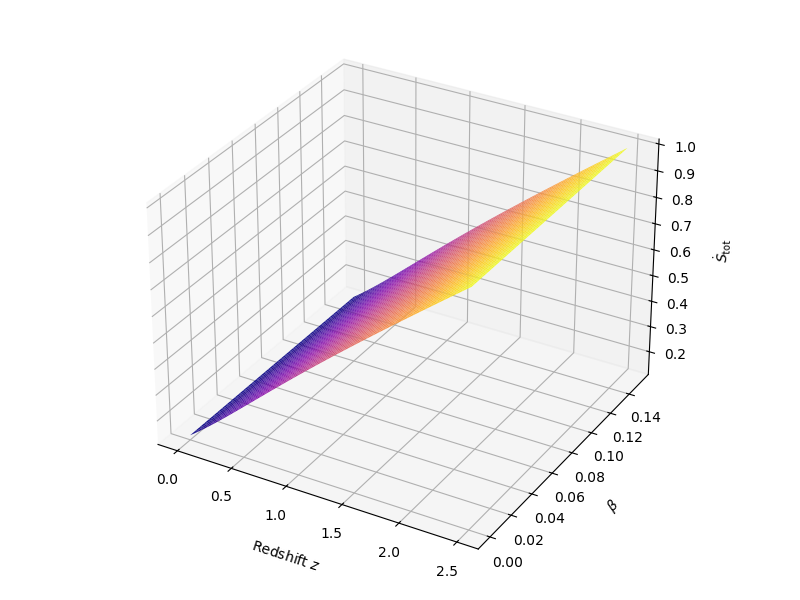}
\caption{$\dot S_{\rm tot}(z,\beta)$}
\label{fig:Sdot_total_beta}
\end{subfigure}

\vspace{0.3cm}

\begin{subfigure}[b]{0.6\linewidth}
\centering
\includegraphics[width=\linewidth]{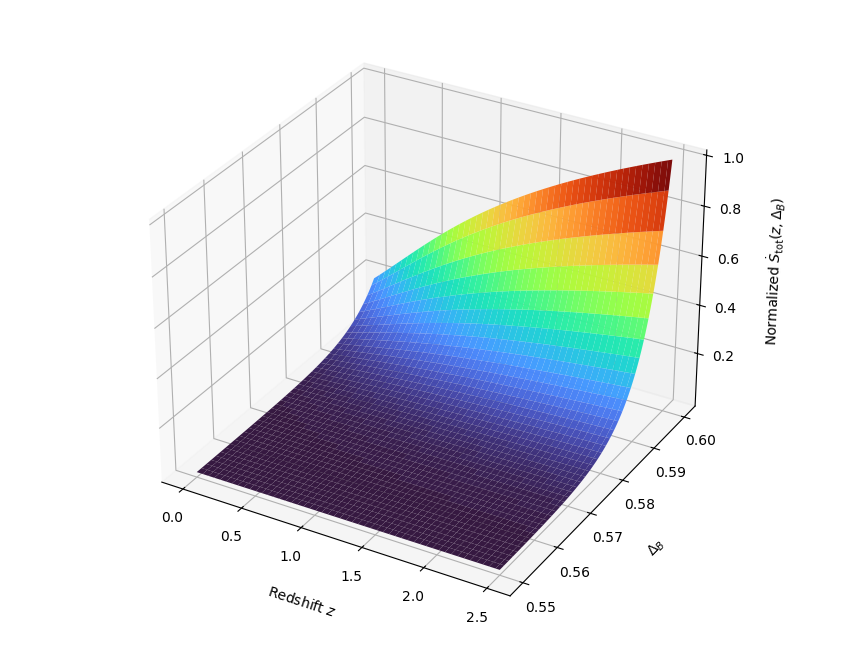}
\caption{$\dot S_{\rm tot}(z,\Delta_B)$}
\label{fig:Sdot_total_DeltaB}
\end{subfigure}

\caption{
Three–dimensional evolution of the normalized total entropy production rate
$\dot S_{\rm tot}$ based on \eqref{Stot_dot_aligned} as a function of the redshift $z$ and the model parameters
$\alpha$, $\beta$, and $\Delta_B$.
All plots are obtained using normalization
$H_0 = 10^{-61} M_{\rm Pl}$ with $\Omega_{m0}=0.3$ and
$\Omega_{r0}=8\times10^{-5}$.
}
\label{fig:Sdot_total_combined}
\end{figure}

From the RHS of Eq.~(\ref{Stot_dot_aligned}) it is apparent that the rate of production of the total entropy depends on the background expansion through \(E(z)\) and its derivative \(dE/dz\). Although the entropy of the cosmic fluid inside the apparent horizon decreases, the Barrow-corrected horizon entropy grows sufficiently fast so that the total entropy variation remains non-negative throughout the cosmic evolution. For the BH--QCDGDE model, it has already been demonstrated that the effective cosmic fluid remains in the quintessence regime, ensuring $\rho_{\rm tot}+p_{\rm tot}>0$, while the expansion of the universe guarantees $H>0$ and $\dot{H}<0$. Consequently, the total entropy variation is expected to satisfy $\dot{S}_{\rm tot}\geq 0$. Thus, the form of the equation makes it apparent that the total entropy variation could satisfy \(\dot S_{\rm tot}\ge 0\), the condition required for the validity of the generalized second law of thermodynamics in the BH--QCDGDE model. However, to further ensure the behavior, we plot $\dot{S}_{tot}$ in Fig.~\ref{fig:Sdot_total_combined}.

Fig.~\ref{fig:Sdot_total_combined} illustrates the evolution of the total entropy production rate $\dot S_{\rm tot}$ as a function of the redshift $z$ and the model parameters $\alpha$, $\beta$, and $\Delta_B$. For all the cases considered here, $\dot S_{\rm tot}$ remains at non-negative level for the entire parameter space. This non-negative behaviour of total entropy production rate confirms the validity of the generalized second law of thermodynamics during the late–time cosmic evolution. Moreover, for the cases of $\alpha$ and $\beta$ a monotone behaviour is observed with $z$. A closer look into the first two panels of Fig.~\ref{fig:Sdot_total_combined}  further reveals that the surfaces have a no significant sensitivity to the parameters $\alpha$ and $\beta$, indicating that the entropy production is not significantly affected by variations in the holographic and ghost dark–energy couplings. Contrary to these two cases, an increasing pattern with the the Barrow deformation parameter $\Delta_B$ leads to a monotonic enhancement of $\dot S_{\rm tot}$. For lower values of $\Delta_B$ it is mostly flattened in the non-negative level. Overall, this reflects the strengthening of the horizon entropy contribution induced by the Barrow correction $\Delta_B$. These results visually presented in Fig.~\ref{fig:Sdot_total_combined}, in general, demonstrate the thermodynamic robustness of the model under consideration and highlight the crucial role of the Barrow entropy deformation in ensuring the generalized second law at late times. It may be noted that in Fig.~\ref{fig:Sdot_total_combined}, $\dot S_{\rm tot}$ has been normalized by its maximum value, since the generalized second law depends only on the sign of the entropy variation. In all cases, $\dot S_{\rm tot}$ remains positive throughout the parameter space, confirming the robustness of the generalized second law of thermodynamics at late times in the BH--QCDGDE framework.

\section{Classical stability analysis}

In cosmological models against small perturbations, the squared speed of sound, denoted here by $v_s^2=\partial p/\partial\rho$ is crucial and has been studied in different literatures. In this context, different schemes of parameterization of the squared speed of sound have been explored in different gravitational frameworks. These include fractal cosmology and nonzero torsion theories, to ensure a stable cosmic evolution \cite{Jawad2019,Rani2022}. More recently, the behavior of the sound speed has also been shown to significantly influence the dynamics of cosmological phase transitions and the resulting stochastic gravitational wave background \cite{Tenkanen2022}. Motivated by these studies, in this study we examine the evolution of $v_s^2$ for the effective cosmic fluid in the present model and identify the regions of the parameter space where $v_s^2\geq 0$, thereby ensuring the absence of classical instabilities throughout the cosmic history. The classical stability of the effective cosmic fluid is examined through the
squared speed of sound, defined as \cite{Jawad2019,Tenkanen2022,Rani2022}
\begin{equation}
v_s^2 \equiv \frac{\dot p_{\rm tot}}{\dot \rho_{\rm tot}},
\label{vsqr}
\end{equation}
where $p_{\rm tot}$ and $\rho_{\rm tot}$ denote the total effective pressure and
energy density, respectively.
A non-negative value of $v_s^2$ ensures stability against small perturbations.

Using the dimensionless Hubble function $E(z)=H(z)/H_0$, in Eqs. \eqref{Ez} and \eqref{eq:dEdz} we have already derived that 
$
E^{2}(z)
=
\Omega_{m0}(1+z)^3
+
\Omega_{r0}(1+z)^4
+
\alpha E^{\,2-\Delta_B}(z)
+
\beta E(z).
$
and $
\frac{dE}{dz}
=
\frac{
3\Omega_{m0}(1+z)^2
+
4\Omega_{r0}(1+z)^3
}{
2E
-
\alpha(2-\Delta_B)E^{\,1-\Delta_B}
-
\beta
}
$. Subsequently, we derive 

\begin{equation}
\begin{aligned}
\frac{d^{2}E}{dz^{2}}
&=
\frac{
6\Omega_{m0}(1+z)
+
12\Omega_{r0}(1+z)^2
}{
\mathcal{D}
}
-
\frac{
\left[
3\Omega_{m0}(1+z)^2
+
4\Omega_{r0}(1+z)^3
\right]
\mathcal{D}'
}{
\mathcal{D}^{2}},
\end{aligned}
\end{equation}
where
\begin{equation}
\mathcal{D}
=
2E
-
\alpha(2-\Delta_B)E^{\,1-\Delta_B}
-
\beta,
\qquad
\mathcal{D}'
=
\left[
2
-
\alpha(2-\Delta_B)(1-\Delta_B)E^{-\Delta_B}
\right]\frac{dE}{dz}.
\label{d2Ddz2}
\end{equation}

The squared speed of sound can then be written solely in terms of $E(z)$ as
\begin{equation}
v_s^2(z)
=
\frac{(1+z)^2 E''(z)}{3(1+z)E'(z)}
-
\frac{(1+z)E'(z)}{3E(z)}.
\label{vsqrz}
\end{equation}
The sign of $v_s^2$ is analyzed numerically to identify stable regions of the
parameter space. Positive values of $v_s^2$ indicate classical stability of the model at late
times, while negative values correspond to unstable regimes.

\begin{figure}[htbp]
\centering

\begin{subfigure}[t]{0.48\textwidth}
\centering
\includegraphics[width=\linewidth]{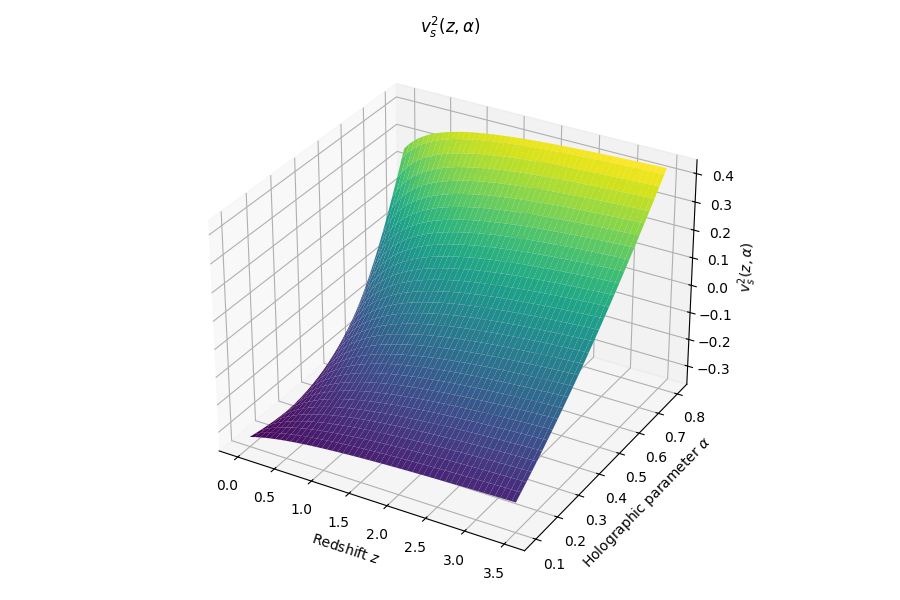}
\caption{
$v_s^2(z,\alpha)$.
}
\label{fig:vs2_alpha}
\end{subfigure}
\hfill
\begin{subfigure}[t]{0.48\textwidth}
\centering
\includegraphics[width=\linewidth]{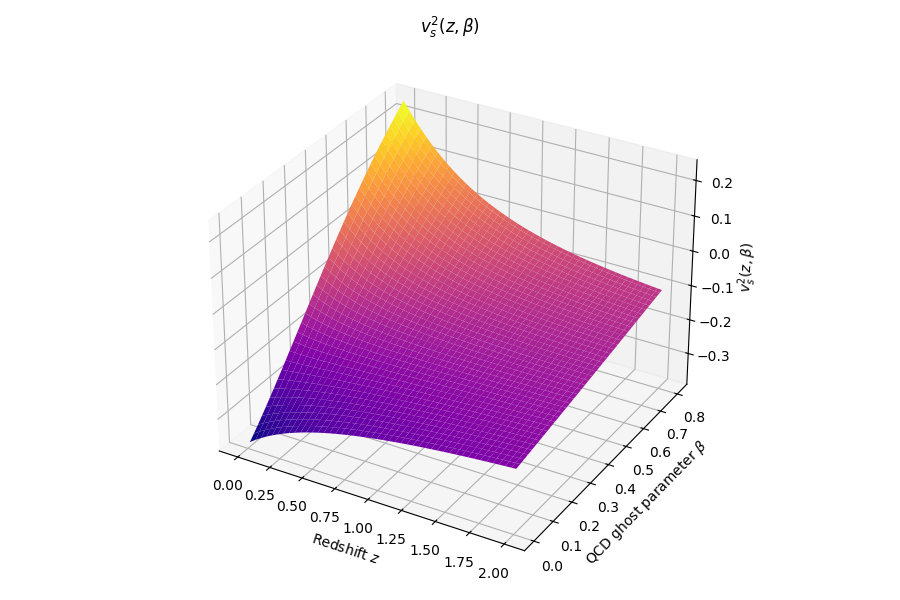}
\caption{
$v_s^2(z,\beta)$.
}
\label{fig:vs2_beta}
\end{subfigure}

\vspace{0.3cm}

\begin{subfigure}[t]{0.6\textwidth}
\centering
\includegraphics[width=\linewidth]{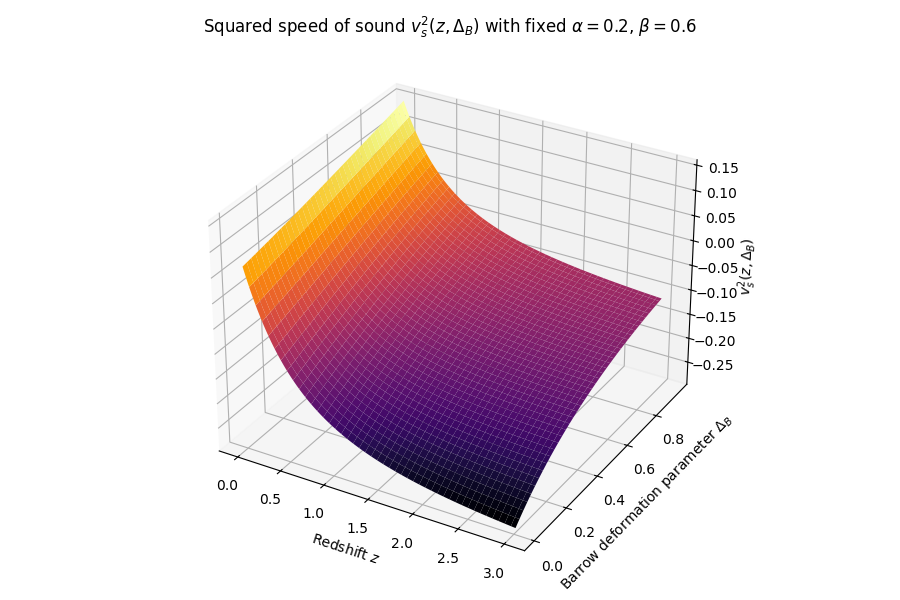}
\caption{
$v_s^2(z,\Delta_B)$.
}
\label{fig:vs2_delta}
\end{subfigure}

\caption{
Three-dimensional behavior of the squared speed of sound
$v_s^2=\dot p_{\rm tot}/\dot\rho_{\rm tot}$ obtained from Eq.~(81).
Panels (a), (b), and (c) illustrate the dependence of $v_s^2$ on the
model parameters $\alpha$, $\beta$, and $\Delta_B$, respectively.
In all cases, the remaining cosmological parameters are fixed to
$\Omega_{m0}=0.3$ and $\Omega_{r0}=8\times10^{-5}$.
Positive values of $v_s^2$ correspond to classically stable regions.
}
\label{fig:vs2_3D}
\end{figure}

Fig.~\ref{fig:vs2_3D} illustrates the three–dimensional evolution of the squared speed of sound $v_s^2=\dot{p}_{\rm tot}/\dot{\rho}_{\rm tot}$, which serves as a diagnostic of the classical stability of the effective cosmic fluid against small perturbations. The left panel shows that increasing the holographic parameter $\alpha$ leads to an increase in the $v_s^2$, particularly at late times ($z\lesssim1$). This behavior indicates an induction of additional curvature due to the Barrow holographic contribution in later stages of the universe. This improves the stability of the effective fluid. The middle panel demonstrates the impact of the QCD ghost parameter $\beta$. For smaller values of $\beta$ the squared speed of sound is negative. However, for larger $\beta$ values the $v_s^2$ becomes non-negative and it further enhances at lower redshifts. This indicates stability in the later epochs of the universe due to ghost contribution on the Hubble evolution. The lower panel corresponds to the role of the Barrow deformation parameter $\Delta_B$. An increase in $\Delta_B$ enhances the entropy–induced corrections to the holographic sector, which in turn amplifies the curvature effects in $E(z)$ and drives $v_s^2$ toward positive values at low redshift. This confirms the influence of the Barrow entropy deformation towards stability of the effective cosmic fluid.

\section{Concluding remarks}
In this work, we have proposed and systematically analyzed a unified dark-energy framework based on the combined effects of Barrow entropy corrections and the QCD ghost mechanism, referred to as the BH--QCDGDE model. The dark-energy density was constructedpresented in Eq.~\eqref{rhoDE} as a hybrid contribution involving a Barrow holographic term proportional to $H^{2-\Delta_B}$ and a linear QCD ghost term proportional to $H$. Eq.~\eqref{rhoDE} therefore incorporates both quantum–gravitational entropy deformations and low-energy QCD vacuum effects within a single phenomenological setup.

We started with the Friedmann equation given in Eq.~\eqref{Ez} and derived the background cosmological evolution and obtained a first-order differential equation for the dimensionless Hubble parameter $E(z)$ [Eq.~\eqref{eq:dEdz}], whose implicit solution is expressed in Eq.~\eqref{eq:implicit}. The resulting expansion history exhibits a smooth transition from a decelerated matter-dominated phase at high redshift to a late-time accelerated phase. This behavior is clearly illustrated through the three-dimensional
evolution of the total equation-of-state parameter $w_{\rm tot}(z)$ and the deceleration parameter $q(z)$ in Figs.~\ref{fig:wtot_3D} and \eqref{fig:q_DeltaB}, obtained from Eqs.~\eqref{w} and \eqref{q}, respectively. Since there was no crossing of phantom bounary and the low sensitivity of the transition epoch to the Barrow deformation parameter $\Delta_B$, we interpreted this as an indication of a viable and observationally consistent evolution of the universe.

We further explored the dependence of the background dynamics on the model parameters by constructing three-dimensional surfaces of $w_{\rm tot}$ and $q$ with respect to $(z,\alpha)$, as shown in Figs.~\ref{fig:wtot_3D_alpha} and \ref{fig:q_3D_alpha}. These figures demonstrate that the Barrow entropy deformation plays a stabilizing role, suppressing unphysical behavior that may arise in the absence of deformation, particularly for large values of the holographic coupling $\alpha$.

An equivalent scalar-field description of the effective dark-energy sector was reconstructed in Sec.~3. Using Eqs.~\eqref{phidotsqr}–\eqref{V_z}, we derived explicit expressions for the scalar-field kinetic term and potential as function of redshift $z$. The reconstructed scalar field exhibits a monotonic evolution, while the potential shows a smooth and well-behaved profile, as depicted in Fig.~\ref{fig:scalar_reconstruction}. This reconstruction confirms that the BH--QCDGDE model admits a quintessence-like scalar-field interpretation without
encountering phantom instabilities.

To gain further analytical insight, we investigated a power-law scale-factor scenario in Sec.~4. The corresponding expressions for the energy density, equation of state, scalar field, and potential [Eqs.~\eqref{rho_pl}–\eqref{V_pl}] reveal that the combined Barrow and QCD ghost contributions naturally lead to accelerated expansion for $n>1$. The reconstructed scalar-field dynamics, as shown in Fig.~\ref{fig:scalar_reconstruction}, further support the consistency of the model across different cosmological realizations.

The generalized second law of thermodynamics was tested in order to thoroughly investigate the thermodynamic viability of the BH--QCDGDE framework. We calculated the production rate of total entropy in Eq.~\eqref{Stot_dot_aligned} using the Barrow-modified horizon entropy [Eq.~\eqref{sh}] and the Gibbs relation for the cosmic fluid [Eq.~\eqref{Th}]. The total entropy variation $\dot S_{\rm tot}$ is non-negative throughout the entire parameter space under consideration, as demonstrated by the three-dimensional plots in Fig.~\ref{fig:Sdot_total_combined}. This confirms the generalized second law at the apparent horizon and highlights the significance of the Barrow entropy correction in preserving thermodynamic consistency.

Finally, we analyzed the classical stability of the effective cosmic fluid through the squared speed of sound, defined in Eqs.~\eqref{vsqr}--\eqref{d2Ddz2} and expressed solely in terms of the Hubble function in Eq.~\eqref{vsqrz}. The three-dimensional surfaces representing the evolving pattern of the squared speed of sound over the parameter spac,e as shown in Fig.~\ref{fig:vs2_3D} reveals that the stability properties are strongly influenced by the interplay between the holographic parameter $\alpha$, the QCD ghost parameter $\beta$, and the Barrow deformation parameter $\Delta_B$. In particular, relatively higher values of $\alpha$ and $\Delta_B$ tend to enhance stability of the model by leading $v_s^2$ toward positive values. On the other hand, significant ghost dominance was found to behave in a different way, but enhancing stability through positive values of $v_s^2$. Such behavior is a generic feature of effective, non-barotropic dark-energy models and does not compromise the viability of the background cosmological evolution.

In conclusion, it may be stated that the BH--QCDGDE model provides a coherent and physically viable description of late-time cosmic acceleration. This suggests that entropy-based holographic corrections and the QCD vacuum effects described in this work have been successfully unified. It is a strong alternative to conventional dark-energy scenarios due to its consistency at the levels of background dynamics, scalar-field reconstruction, thermodynamics, and classical stability. This framework may be expanded in subsequent studies to incorporate perturbation growth, observational constraints from large-scale structure, and potential embeddings within modified gravity theories like $f(T)$ or $f(Q)$ gravity.

\section*{Acknowledgment}
This research was funded by the Science Committee of the Ministry of Science and Higher Education of the Republic of Kazakhstan (Grant No. AP32720463).

\end{document}